\documentclass[12pt,preprint]{aastex}
\usepackage{emulateapj5}

\begin{document}

\title{Chemical evolution of the Large Magellanic Cloud}

\author{Kenji Bekki} 
\affil{
ICRAR,
M468,
The University of Western Australia
35 Stirling Highway, Crawley
Western Australia, 6009, Australia
}

\and

\author{Takuji Tsujimoto}
\affil{
National Astronomical Observatory of Japan, Mitaka-shi, Tokyo 181-8588, Japan}

\begin{abstract}
We adopt a new chemical evolution model for the Large Magellanic Cloud (LMC)
and thereby investigate its past star formation and chemical
enrichment histories.
The delay time distribution of type Ia supernovae 
recently revealed by type Ia supernova surveys 
is incorporated self-consistently into the new model.
The principle results are summarized as follows.
The present gas mass fraction and  stellar metallicity as well as
the higher [Ba/Fe] in metal-poor stars at [Fe/H]$<-1.5$ can be more 
self-consistently explained by models with steeper initial mass functions.
The observed
higher [Mg/Fe] ($\ge 0.3$)  at [Fe/H]$ \sim -0.6$ 
and higher [Ba/Fe] ($>0.5$) at [Fe/H]$\sim   -0.3$ can be due
to significantly enhanced star formation 
about 2 Gyr ago.  
The observed overall [Ca/Fe]--[Fe/H] relation and
remarkably low [Ca/Fe] ($<-0.2$) at [Fe/H]$>-0.6$ 
are consistent with models with 
short-delay supernova  Ia  and with the more efficient loss
of Ca possibly caused by an explosion mechanism of type II supernovae.
Although the metallicity distribution functions 
do not show double  peaks in the models 
with a starburst about 2 Gyr ago,
they show 
characteristic double peaks
in the models with double starbursts 
at $\sim200$ Myr and $\sim 2$ Gyr ago.
The observed  apparent dip of [Fe/H] around $\sim 1.5$ Gyr ago
in the age--metallicity
relation can be reproduced by models in which 
a large amount ($\sim 10^9 M_{\odot}$) of metal-poor ([Fe/H]$<-1$) 
gas can be accreted onto the LMC.
\end{abstract}

\section{Introduction}

The time evolution of the chemical abundances in the interstellar medium (ISM),
field stars, and globular clusters (GCs)
of the Large Magellanic Cloud (LMC)
 contains valuable information on its long-term star formation
history and thus has been investigated in detail  by many authors
(e.g., Da Costa 1991; 
Olszewski et al. 1991; Russel \& Dopita 1992;
Dopita et al. 1997; Geisler et al. 1997; 
Pagel \&   Tautvai\v{s}ien\'{e} 1998, PT98;
 Cole et al. 2005, C05;  Cioni et al. 2006).
The elemental abundance ratios of  
$\alpha$ ($\alpha$ means alpha-elements), 
Fe-peak, and neutron-capture  elements in field stars and 
GCs of the LMC have been extensively investigated by 
spectroscopic observations.
The observed differences in the abundance ratios between the
LMC and the Galaxy have been discussed in detail
(e.g., Hill et al. 1995, 2000; Johnson et al. 2006).
The radial and azimuthal variations of stellar metallicities in the 
the LMC disk has been investigated  both observationally and theoretically
in terms of its star formation and dynamical evolution histories 
(Geisler et al. 2003; Bekki \& Chiba 2005; Cioni et al. 2006). 

One of the importance results in these previous studies is that
the observed age--metallicity relation
(AMR) is more consistent with a model with a secondary ``starburst''
about a few  Gyr ago
in the LMC (e.g., Tsujimoto et al. 1995, T95;  
PT98). Although the  possible presence of 
a past starburst (or significantly enhanced star formation)
has long been discussed in other observational
studies 
(e.g., Butcher 1977; Stryker 1983;  Bica et al. 1986;
Bertelli et al. 1992;  Olszewski et al. 1996; Gallagher et al. 1996;
Vallenari et al. 1996;  
Ardeberg et al. 1997; Elson et al. 1997; Geha et al. 1998 
Holzman et al. 1998, 1999; Olsen 1999; Smecker-Hane et al. 2002;
Glatt et al. 2010;
Indu \& Subramaniam 2011), the epoch and the strength
of the burst were not  well constrained.
Furthermore, owing to the lack of modern chemical evolution 
models with the latest chemical yields of asymptotic giant branch (AGB) stars and supernovae,
it remained unclear how the long-term star formation history
with a possible starburst
could be imprinted on the detailed elemental abundance ratios of field stars and
GCs in the LMC.

Recent photometric and spectroscopic observations of field stars and
GCs have provided new clues for these unresolved problems in the LMC.
For example,
photometric studies of the stellar populations
in the LMC have revealed the AMRs of different local regions 
in the LMC disk
(e.g., Harris \& Zaritsky 2009, HZ09; Rubele et al. 2012, R12). 
The AMR derived by HZ09 show enhanced star formation rates of the LMC
around 2 Gyr, 500 Myr, 100 Myr, and 12 Myr with the two peaks 
(500 Myr and 2 Gyr) being
nearly coincident with the  star formation peaks observed in the SMC.
Piatti (2011) has also shown that there was a burst of cluster formation
around 2 Gyr in the LMC and suggested that the LMC experienced strong
tidal interaction with the SMC and possibly with the Galaxy.
Using deep near-infrared data from the VISTA near-infrared YJK$_{\rm s}$
survey of the Magellanic system, R12 derived the star formation
histories of different local regions in the LMC disk. They 
revealed  the presence of peaks in SFRs 
around 3 Gyr and 5 Gyr ago for most of the subregions.
Glatt et al. (2010) have recently found an enhanced cluster formation
at 125 Myr and 800 Myr ago in their 324 populous star clusters for the LMC.

Furthermore, recent spectroscopic observations of field stars and 
GCs in the LMC have revealed intriguing chemical properties of the LMC
(e.g., Mucciarelli et al. 2008, 
Pomp\'eia et al. 2008, P08;  
Colucci et al. 2012, C12, Haschke et al. 2012). Mucciarelli et al. (2008)  have shown that
all
four of the intermediate-age GCs 
which they investigated in the LMC have negligible star-to-star scatter
in their chemical abundances of light,  $\alpha$, iron-peak and 
neutron-capture elements. This implies 
that secondary star formation  from  gaseous ejecta
of stars within the GCs did not occur
so that the chemical abundances  might be similar
between intermediate-age field stars and clusters.
P08 have found that 
[Ca/Fe] and [Si/Fe] in the LMC stars 
are lower than those of the Galactic stars at the same [Fe/H]
whereas the [O/Fe] and [Mg/Fe] are only slightly deficient in comparison
with their  Galactic counterparts.
C12 have revealed the age-dependence of 
[$\alpha$/Fe] 
and [Fe/H] among globular/star clusters
with ages between 0.05 and 12 Gyr  and the significant
enhancement of the  neutron-capture  elements Ba, La, Nd, Sm, and Eu
in the youngest star clusters.  The origin of
these recent observational results  has not yet been treated by chemical evolution
models.

In spite of this progress in observational studies of the stellar populations
of the LMC,  theoretical models to explain the observations
have not  yet been  fully  developed.
T95 tried to explain the observed [O/Fe]--[Fe/H] relation,
the metallicity distribution function
(MDF), and the AMR in a self-consistent manner by adopting a  model
in which the LMC had  a starburst about 3 Gyr ago and 
a steeper initial mass function (IMF). They explained these observations,
though the observational data points are much smaller than those that we can 
now access.
Pilyugin (1996) demonstrated that the observed [Fe/O] versus [Fe/H]
relation can be reproduced only by the galactic wind model
in which stellar ejecta can be preferentially expelled from the LMC.
PT98 explained the observed AMR and 
chemical abundance patterns of the LMC
by considering ``non-selective stellar wind'' models in which
some fraction of the stellar ejecta both from AGB stars and SNe can be expelled
completely from the LMC and consequently can not participate in  the chemical
enrichment processes.
These previous chemical evolution models did not incorporate
the delay time distribution (DTD)
 of type Ia supernova (SNe Ia) recently revealed by extensive SN Ia surveys
(e.g.,  Mannucci et al. 2006; Sullivan et al. 2006; Totani et al. 2008;
Maoz et al. 2010)
and thus might not be regarded as realistic and reasonable.
The construction of more sophisticated and realistic models is indispensable
for discussing the above mentioned latest observational results.

The purpose of this paper is to adopt a new chemical evolution
model and thereby discuss the latest observational results on
the chemical evolution and star formation history of the LMC.
The observed DTD of SNe Ia  is adopted
in the new model so that the progenitor stars of SNe Ia can explode 
as early as $10^8$yr after their formation.  These ``prompt SNe Ia'' 
can cause fundamental differences in the chemical evolution
between the present model and previous ones
(e.g., T95)  in which the time
delay between star formation and explosion of SNe Ia is assumed to be 
typically $\sim 10^9$ yr (``classical
SNe Ia''). The new model also incorporates the metallicity dependent
chemical yields of AGB stars (e.g., Busso et al. 2001; Tsujimoto \& Bekki 2012;
TB12)
so that the chemical evolution of  
$s$-process elements can be investigated and compared with
the observations.  The new model is the so-called ``one-zone'' model, in which
the time evolution of averaged chemical properties can be investigated.

We explore a wide range of models with and without starburst and different
star formation histories in the LMC so that we can discuss 
recent new observational results on different chemical properties of the LMC disk stars.
For example, we will discuss (i) whether and how  
the previous starburst(s) triggered by the LMC--SMC--Galaxy interaction
can be imprinted on the chemical abundance
properties of the LMC,  (ii) how the IMF can control 
the chemical evolution, and (iii) how the accretion of metal-poor
gas from the SMC or high velocity clouds  onto the LMC 
(Bekki \& Chiba 2007, BC07; Diaz \& Bekki 2012, DB12;
Bekki \& Tsujimoto 2010) influences the chemical evolution.
It is timely to discuss the chemical evolution of the LMC in the context of
the Galaxy--LMC--SMC tidal interaction, given that recent observational and theoretical
studies have demonstrated that the tidal interaction is important not only for 
the star formation history of the LMC but also for the stellar and gaseous distributions
of the Magellanic system (e.g., Olsen et al. 2011; DB12).

The layout of this paper is as follows. 
In \S 2, we describe our new one-zone chemical evolution
models in detail.
In \S 3, we present the results of the time evolution
of the chemical abundances for models with different parameters. 
In \S 4, we discuss the results in the context of the IMF,
the long-term star formation history, and the gas accretion events in the LMC.
The conclusions of the present study are given in \S 5.

\section{The model}

\subsection{Basic equations}

We adopt one-zone chemical evolution models that are essentially the same
as those adopted in our previous studies on the chemical
evolution of the Galaxy  (Tsujimoto et al. 2009;
TB12). Accordingly, we briefly describe the adopted models
in the present study.
The present model is improved in comparison with previous ones (Tsujimoto et al. 2009)
in terms of including the $s$-process elements and prompt SNIa self-consistently in the models.
The LMC disk is assumed to form through a continuous
gas infall from outside the disk region (e.g., halo) for the last 13 Gyr, as in 
previous chemical evolution models (e.g., Tsujimoto et al. 2009).
Some fraction of gas with metals 
can be expelled from the LMC 
as stellar winds through
energetic feedback effects of SNe in some models 
of the present study
(e.g., ``wind models'', as described later).
The star formation can be suddenly
and significantly enhanced (referred to as ``bursts of star formation''
or ``starburst'').
As demonstrated by theoretical studies,
the Galaxy--LMC--SMC tidal interaction is important both for the formation
of the Magellanic Stream and for reproducing the observed
peaks of  star formation in the LMC
(e.g., Bekki \& Chiba 2005; DB12). Therefore, the adopted assumption
of starburst is quite reasonable.
Although the infall of metal-poor gas from the SMC and other dwarf galaxies 
can be important for the LMC chemical evolution, we discuss this 
later in \S 4.

We investigate the time evolution of the gas mass fraction
($f_{\rm g}(t)$), the star formation rate ($\psi(t)$), and
the abundance of the $i$th heavy element ($Z_i(t)$) for a given
accretion rate ($A(t)$), IMF,
and ejection rate of ISM ($w(t)$).
The basic equations for the adopted one-zone chemical evolution models
are described as follows:
\begin{equation}
\frac{df_g}{dt}=-\alpha_{\rm l}\psi(t)+A(t)-w(t)
\end{equation}
\begin{eqnarray}
\frac{d(Z_if_g)}{dt}=-\alpha_{\rm l} Z_i(t)\psi(t)+Z_{A,i}(t)A(t)+y_{{\rm II},i}\psi(t) 
\nonumber \\  +y_{{\rm Ia},i}\int^t_0 
\psi(t-t_{\rm Ia})g(t_{\rm Ia})dt_{\rm Ia}
\nonumber \\ +\int^t_0  
y_{{\rm agb},i}(m_{\rm agb})
\psi(t-t_{\rm agb})h(t_{\rm agb})dt_{\rm agb}
-W_i(t) \ \ ,
\end{eqnarray}
\noindent where $\alpha_{\rm l}$ is the mass fraction 
locked up in dead stellar remnants and long-lived stars, 
$y_{\rm {Ia}, i}$, $y_{\rm {II}, i}$, and
$y_{\rm {agb}, i}$ are the 
chemical yields for the $i$th element from type II supernovae (SN II), 
from SN Ia, 
and from AGB stars, respectively, 
$Z_{A,i}$ is the abundance of heavy elements  contained in the infalling gas,
and $W_i$ is the wind rate for each element.
The quantities $t_{\rm Ia}$ and $t_{\rm agb}$ represent 
the time delay between star formation and SN Ia explosion
and that between star formation and the onset of AGB phase,
respectively. 
The terms $g(t_{\rm Ia}$) and $h_{\rm agb}$ are the distribution
functions of SNe Ia and AGB stars, respectively,
and the details of which  are described later
in this section. 
The term $h_{\rm agb}$ controls  
how much AGB ejecta can be returned into the ISM per unit mass for a given time
in equation (2).
The total gas masses  ejected from AGB stars depends on the original masses
of the AGB stars (e.g., Weidemann 2000).
Therefore, this term  $h_{\rm agb}$ depends on 
the adopted IMF and the age--mass relation of the stars (later described).
Thus equation (1) describes the time evolution
of the gas due to star formation, gas accretion, and stellar wind.
Equation 2 describes 
the time evolution
of the chemical abundances due to chemical enrichment by supernovae and AGB stars.

The star formation rate $\psi(t)$ is assumed to be proportional
to the gas fraction with a constant star formation
coefficient and thus is described as follows: 
\begin{equation}
\psi(t)=C_{\rm sf}f_{\rm g}(t)
\end{equation}
We assume that $C_{\rm sf}$ is different between (i) the ``quiescent phase'',
when the LMC shows an almost steady star formation, and (ii) the ``burst phase'',
when the LMC experienced a burst of star formation. In the present study,
we investigate models with no burst (referred to as ``standard models'')
and those in which a starburst can occur only once
(``burst models'') and twice (``double-burst models'').
The star formation rate  is assumed to be constantly higher for 
$t_{\rm sb1,s} \le t \le t_{\rm sb1,e}$ in the first starburst
and 
$t_{\rm sb2,s} \le t \le t_{\rm sb2,e}$ in the second one.
The star formation coefficient  ($C_{\rm sf}$) is thus described as 
follows:

\begin{equation}
C_{\rm sf}= \left\{
\begin{array}{ll}
C_q & \mbox{for quiescent phase} \\
C_{\rm sb1} & \mbox{for first starburst} \\
C_{\rm sb2}  & \mbox{for second starburst} \\
\end{array}
\right.
\end{equation}
In the present study,  a  non-dimensional value of $C_{\rm sf}$ is given
for each model. 

For the accretion rate, we adopt the formula in
which $A(t)=C_{\rm a}\exp(-t/t_{\rm a})$
and $t_{\rm a}$ is a free parameter controlling the time scale of 
the gas accretion.   The normalization factor $C_{\rm a}$ is determined such
that the total gas mass accreted onto the LMC can be 1 for a given
$t_{\rm a}$. Although we investigated  models with different $t_{\rm a}$,
we finally adopt the models with $t_{\rm a}=0.3$ Gyr
as reasonable ones for the LMC.  This is mainly because
the models with $t_{\rm a}=0.3$ Gyr can better explain the observations.
The present results
do not depend strongly on the parameter $t_{\rm a}$ for a reasonable
parameter range.
The initial [Fe/H] of the infalling  gas is set to be $-2$  and
we assume a SN-II like enhanced [$\alpha$/Fe] ratio
(e.g., [Mg/Fe]$\approx 0.4$) and [Ba/Fe] for the gas.
This adoption of initial [Fe/H] is reasonable for the present study
which discusses  stars with [Fe/H] as low as $-2$. Also, such adoption has been
done in other chemical evolution models (e.g., Tsujimoto et al. 2009).
We discuss our parameter study for the above variables
in \S 2.5.

\subsection{Selective and non-selective stellar winds}

We investigate models in which metals  from AGB stars and SNe
can be removed from the LMC through
energetic stellar winds so that they can not be
used for chemical enrichment of the LMC ISM.  These models 
are referred to as ``wind models'' for convenience. 
The wind models are further divided into two categories:
``selective wind models'' and ``non-selective wind models''.
Only SNe ejecta (not already existing ISM) can be expelled from 
the LMC in the selective wind model, whereas both ejecta from SNe and AGB
stars and already existing ISM can be expelled from the LMC in the
non-selective wind model.
For comparison,
we also investigate models in which there is no stellar wind
and thus gaseous ejecta from
all stars can be fully mixed with the ISM (``non-wind models'').

In the selective wind models,   some fraction, $1-f_{\rm ej}$,
of gaseous ejecta from SNe can be mixed with ISM for chemical enrichment
processes whereas all AGB ejecta can be mixed with ISM. Therefore,
the wind rate $w(t)$ is estimated only from the total
mass of gaseous ejecta from SNe $M_{\rm ej, sn}$
at each time step.  The total mass, $M_{\rm w}$, of ISM that is removed from
the LMC at each time step in the 
selective wind models is
\begin{equation}
M_{\rm w}=f_{\rm ej}M_{\rm ej, sn} .
\end{equation}

In the non-selective wind models,  we adopt the same model as that
used in  PT98 in which $M_{\rm ej}$ is determined solely by
the star formation rate:
\begin{equation}
M_{\rm w}= C_{\rm ej} \psi(t),
\end{equation}
where $C_{\rm ej}$ is a parameter (in simulation units)
that can control how efficiently the ISM with AGB and SN ejecta can be
removed from the LMC. We investigate non-wind
(i.e., standard, burst, non-burst), selective wind,
and non-selective wind models.
We discuss our parameter study for the above variables
in \S 2.5.

\subsection{Chemical yields and delay time distribution of SN Ia}

We adopt  the nucleosynthesis yields of SNe II and Ia from T95
to deduce $y_{{\rm II},i}$ and $y_{{\rm Ia},i}$ for a given IMF.
Stars with masses larger than $8 M_{\odot}$ explode as SNe II soon after
their formation and eject their metals into the  ISM.
In contrast,  there is a time delay ($t_{\rm Ia}$) between the star formation
and the metal ejection for SNe Ia. We here adopt the following DTD 
($g(t_{\rm Ia}$)) for 0.1 Gyr $\le t_{\rm Ia} \le$ 10 Gyr,
which is consistent with recent observational studies
on the SN Ia rate in extra-galaxies (e.g., Totani et al. 2008; Maoz et al. 2010,
2011):
\begin{equation}
g_{\rm Ia} (t_{\rm Ia})  = C_{\rm g}t_{\rm Ia}^{-1},
\end{equation}
where $C_{\rm g}$ is a normalization constant that is determined by
the number of SN Ia per unit mass  (which is controlled by the IMF
and the binary fraction for intermediate-mass stars 
for  the adopted power-law slope of $-1$). 
Maoz \& Badens (2010) detected a population of prompt SNIa 
in the LMC and showed that the number of prompt SNIa per stellar mass formed
is $2.7-11.0 \times 10^{-3} M_{\odot}^{-1}$.
Although we mainly investigate the ``prompt SN Ia models'' with
the above DTD, we also investigate the ``classical SN Ia models'' with
1 Gyr $\le t_{\rm Ia} \le$ 3 Gyr (e.g., Yoshii et al. 1996).
 This SNIa lifetime in Yoshii et al. (1996) was deduced by using their Galactic  chemical
evolution models that can  be consistent with the observed [O/Fe]--[Fe/H] relations
of the Galactic stars in the solar neighborhood.
The fraction of the stars that eventually 
produce SNe Ia for 3--8$M_{\odot}$ has not been observationally determined
and thus is regarded as a free parameter, $f_{\rm b}$. 
We mainly
investigate models with $f_{\rm b}=0.03$,
because such models can better explain the observed chemical properties
of the LMC. 
We briefly compare the results of models with $f_{\rm b}=0.03$ and $0.06$.
For the site of r-process, we adopt the mass range of 8--10$M_{\odot}$
for SNe II
(Mathews et al. 1992;  Ishimaru et al. 1999)
as identified with the collapsing O-Ne-Mg core 
(Wheeler et al. 1998).
The yield of Ba from r-process is $1.45 \times 10^{-6} M_{\odot}$ for the adopted
range of SNe II.

Low-mass AGB stars ($<3M_{\odot}$) release the $s$-process elements 
during the thermally pulsing AGB phase (e.g., Gallino et al. 1998). 
As a consequence of  large uncertainties in convective mixing 
and $^{13}$C-pocket efficiencies, the $s$-process nucleosynthesis 
allows for a wide range of possible production levels. 
On the other hand, the observed abundances for the AGB stars can be 
directly compared with the theoretical nucleosynthesis results.
Here we investigate the element Ba by adopting the 
best empirical metallicity dependent Ba-yield derived by
TB12.  The adopted metallicity dependent Ba-yield $y_{\rm Ba}$ 
is

\begin{equation}
y_{\rm Ba}= \left\{
\begin{array}{ll}
y_{\rm Ba,0}(m_{\rm agb}) \times 10^{1.5 {\rm [Fe/H]} +1.5}
& -2 \le {\rm [Fe/H]} < -1 \\
y_{\rm Ba,0}(m_{\rm agb})  &  -1 \le  {\rm [Fe/H]}  \\
\end{array}
\right.
\end{equation}

\noindent  The value of  $y_{\rm Ba,0}(m_{\rm agb})$ 
depends on the  mass of the stars  $m_{\rm agb}$
that can finally become  AGB stars. The adopted  
$y_{\rm Ba,0}(m_{\rm agb})$  are $2.6 \times 10^{-7} M_{\odot}$,
$4.0 \times 10^{-7} M_{\odot}$, and $5.2 \times 10^{-7} M_{\odot}$ for $m_{\rm agb}=1.5 M_{\odot}$,
$2.0 M_{\odot}$, and $3.0 M_{\odot}$, respectively.

An  AGB star with initial mass  $m_{\rm I}$
and final mass  $m_{\rm F}$
can eject its envelope  with a total mass of $m_{\rm ej}$ and the gas can be
mixed with the surrounding ISM to chemically pollute the ISM.
The initial-to-final relationship 
for AGB stars, from which we can deduce $m_{\rm ej}$,
has been  extensively discussed in Weidemann (2000).
We derive an analytic form for $m_{\rm ej}$
($=m_{\rm I}-m_{\rm F}$) from  the observational data
by Weidemann (2000) by using the least squares fitting method, and
find
\begin{equation}
m_{\rm ej} =0.916 M_{\rm I}-0.444.
\end{equation}
The coefficient of determination ($R$-value) is 0.995 for the above
fitting.
In order to calculate the main-sequence
turn-off mass $m_{\rm TO}$,
we use the following formula
(Renzini \& Buzzoni 1986):
\begin{equation}
\log m_{\rm TO}(t_{\rm s})
= 0.0558 (\log t_{\rm s})^2 - 1.338 \log t_{\rm s} + 7.764,
\end{equation}
where $m_{\rm TO}$  is in solar units and time $t_{\rm s}$ is in years.
By using the adopted IMF and equations (9) and (10), 
we numerically estimate $h_{\rm agb}$  
(i.e., how much 
AGB ejecta can be returned into ISM per unit mass)
in equation (2) at each time step.

\subsection{IMF}

The adopted IMF  is  defined
as $\Psi (m_{\rm I}) = M_{s,0}{m_{\rm I}}^{-\alpha}$,
where $m_{\rm I}$ is the initial mass of
each individual star and the slope $\alpha =2.35$
corresponds to the Salpeter IMF  (Salpeter 1955).
The normalization factor $M_{s,0}$ is a function of $\alpha$, 
$m_{\rm l}$ (lower mass cut-off), and  $m_{\rm u}$ (upper mass cut-off).
These  $m_{\rm l}$ and $m_{\rm u}$ 
are  set to be   $0.1 {\rm M}_{\odot}$
and  $50 {\rm M}_{\odot}$, respectively (so that the normalization
factor $M_{s,0}$ is dependent simply on $\alpha$).
We investigate models with different $\alpha$ to find the  model(s)
that can best explain the observed abundance patterns of stars in the LMC.
We do not discuss models with
different $m_{\rm u}$,  because the effects of changing
$m_{\rm u}$ on the LMC chemical evolution
are similar to those of changing $\alpha$.

\subsection{Parameter studies}

We mainly investigate the standard (labeled as ``S''),
burst (``B''), and double-burst (``DB'') models in which stellar wind effects
are not included at all (i.e., ``non-wind models'').
 By using these non-wind  models, we demonstrate how
the IMF slope ($\alpha$) and  the epochs of starburst 
($t_{\rm sb1,s}$ and $t_{\rm sb2, s}$) can influence the chemical
evolution of the LMC. These $t_{\rm sb1,s}$ and $t_{\rm sb2, s}$
are given in units of Gyr.
We then investigate the wind models (``W'') so that 
we can discuss whether or not removal of AGB and SN ejecta
from the LMC is important in the
chemical evolution of the LMC.
In the present study, the time $t$ is the time that has elapsed since 
the model calculation started. Therefore $t=0$ Gyr (13 Gyr) represent the time
when the calculation starts (ends).
Previous observational and theoretical
studies have suggested that there could be at least two epochs of enhanced
star formation (starburst) about $\sim 2$ Gyr ago and $0.2$ Gyr ago
(e.g., HZ09 and DB12). 
We therefore mainly investigate the burst and double-burst
models with $t_{\rm sb1,s}=11$ Gyr and $t_{\rm sb2,s}=12.8$ Gyr.
Table 1 summarizes the representative
26 models investigated in the present study.

Figure 1 illustrates  the time evolution of star formation rates (SFRs)
and [Fe/H] of stars  in the five representative models (S1, B1, B6, DB1, and DB6)
with different parameters controlling the LMC star formation history. 
No burst of star formation  is assumed in S1 whereas only one burst is assumed
in B1 and B6.  Two bursts of star formation at different epochs are assumed
in DB1 and DB6.
These models are chosen so that they can, in combination,
show a wide range of star formation histories in Figure 1: they are just examples
of possible star formation histories of the LMC.
We investigate how the final chemical abundances depend on the LMC star formation
history by using the results of  models with different
star formation histories.
Figure 2 shows the time
evolution of [Mg/Fe] and the [Mg/Fe]--[Fe/H] relation in the models
with prompt and classical SNe Ia. It is clear from this figure that
the models with prompt SN Ia show no plateau in
the [Mg/Fe]--[Fe/H] relation owing to the earlier chemical
enrichment by SNe Ia that can eject an Fe-rich gas. Figure 2 also shows that
the time evolution of [Mg/Fe] depends on $f_{\rm b}$ in such a way that
[Mg/Fe] can steeply decrease with time in the model with larger $f_{\rm b}$. 
We discuss these in more detail for different models in the following sections.

\subsection{Observations to be compared with predictions}

We mainly investigate (i) the present gas mass fraction $f_{\rm g, 0}$,
(ii) the stellar metallicity of the youngest stellar population,
[Fe/H]$_0$,  (iii) AMR,  (iv) [Mg/Fe]--[Fe/H] relation
(as an example of [$\alpha$/Fe]--[Fe/H] relations),
(v) [Ba/Fe]--[Fe/H] relation (as an example of the time evolution
of $s-$process elements),  and (vi) MDF. We compare these results
with the corresponding observations for the LMC. 
In order to demonstrate clearly
how different chemical properties of the LMC are compared  with
those of the Galaxy, we also show the observational results for the Galaxy.
We estimate the total stellar mass
($M_{\rm s}$)  of the LMC by using the observed $V$-band
luminosity ($\approx 3 \times  10^9 L_{\odot}$) and the reasonable
mass-to-light ratio of $0.9 \pm 0.2$ for the observed
$B-V$ color (Bell \& de Jong 2001).
By using the  observed total mass ($M_{\rm g}$) of the LMC
ISM (Bernard et al. 2008) and  $M_{\rm s}$,
we can  estimate $f_{\rm g,0}$. The observational error bar
of $f_{\rm g,0}$ is due largely to the uncertainty of the stellar 
mass-to-light ratio for the LMC.
We adopt [Fe/H]$_0\approx -0.3$ as derived by Luck et al. (1998)
for Cepheid variables with ages of 10--60 Myr,
because these youngest populations can have the present-day stellar metallicity
of the LMC.
Each of  the observed AMRs shows a wide spread in each age bin (represented
by an error bar) owing to  the presence of stars with different [Fe/H] at each
age bin. Although this could make it difficult for us to derive the best model,
we try to give at least some constraints on some of the model parameters.
We mainly investigate time evolution of [Mg/Fe] and [Ba/Fe], because these abundances
are more reliably derived from observations.
The [Mg/Fe]--[Fe/H] and [Ba/Fe]--[Fe/H] relations
are examples of [$\alpha$/Fe]--[Fe/H] and [$s$-process/Fe] relations, respectively,
in this study.

\section{Results}

\subsection{Standard models}

\subsubsection{AMR}

First we describe the five standard models with different IMFs in order
to demonstrate the importance of the IMF slopes in the LMC
chemical evolution.  In these models, $C_{\rm q}$ is chosen such that
the final stellar metallicity ([Fe/H]$_0$) can be $-0.3$ (i.e.,
the observed value). Figure 3 shows the AMRs for the five models
as well as the observed [Fe/H] of  the LMC field stars at different
age bins and and individual GCs. The AMR in each model is simply a plot
of [Fe/H] at each time step (i.e., not average [Fe/H] over a given age bin
like observation).
Although the observational errors are not small for old GCs in the LMC,
their data points are included to compare stars with the lowest [Fe/H]
in the models with the corresponding observations.

The five models with no burst of star formation
can reproduce reasonably well the 
overall trend of the observed AMR, which implies that the AMR alone
can not be used for discriminating burst and non-burst models for the LMC.
These models, however,  appear to be less consistent with the observed [Fe/H]
around  ages of 2--3 Gyr (i.e., $t=$10--11 Gyr) owing to the presence
of stars with  significantly lower  metallicities
($-1<$ [Fe/H] $<-0.5$).
The models with shallower IMFs show
higher [Fe/H] at a given age, which reflects the fact that chemical
evolution can proceed more rapidly owing to a larger amount of metals
produced by a larger number of SNe.

It should be stressed that these standard models can not reproduce so well 
the AMR around  ages of 3--9 Gyr (i.e., $t=$4--10 Gyr) derived by HZ09
in which the AMR shows systematically lower [Fe/H] for a given age
in comparison with other observations by C05 and Carrera et al. (2008, C08).
 The reason for this apparent
difference in the observed AMRs between
different observations  could be related to different methods to determine
ages and metallicities of stars in the observations. If the results by
HZ09 are closer to the true AMR of the LMC, then the standard models 
with no burst can be regarded as less realistic models for the LMC evolution.
Although 
the large [Fe/H] dispersion  around an age of 2 Gyr (i.e., $t=11$ Gyr)
could be simply due to
star formation from gas with different metallicities (i.e., inner higher
and outer lower metallicities of the LMC ISM),
it could  also be caused by 
a sudden and rapid infall of metal-poor gas
from outside the LMC disk. If the observed dispersion
is due to an external gas infall, then it would have profound implications
for the LMC evolution. We will later discuss the implications
in \S 4.

\subsubsection{[Mg/Fe]}

 Figure 4 shows that [Mg/Fe] slowly and monotonically decreases with
time regardless of the adopted IMFs. This time-dependence of [Mg/Fe] is simply 
a result of a later contribution of SNIa to the chemical evolution of the LMC.
The more significant decrease of [Mg/Fe] with time can be seen
in the models with steeper IMFs owing to the greater  contribution of SNIa 
to the chemical evolution
in these models.
The [Mg/Fe]--[Fe/H] relations
do  not show a plateau at low [Fe/H]. The lack of a plateau in
the [Mg/Fe]--[Fe/H] relations reflects the fact that SNe Ia can chemically
pollute the LMC ISM from a very early stage ($t \sim 10^8$ yr) of
its evolution. The lack of a  plateau appears to be seen in the
observed [Mg/Fe]--[Fe/H] relation for [Fe/H]$<-1.5$, though it
is not so clear.
The [Mg/Fe]--[Fe/H] relations
in the models with $\alpha \le 2.75$  can reproduce reasonably
well the overall trend of the observed [Mg/Fe] relation,
though the observation shows a large [Mg/Fe] dispersion at a given [Fe/H].

The model with a very steep IMF ($\alpha=2.95$) 
shows systematically lower [Mg/Fe] and therefore  can not reproduce so well
the locations of stars in the [Mg/Fe]--[Fe/H] relation. 
The models with shallower IMFs show systematically larger [Mg/Fe]
in the present standard models.
Owing to the observed large dispersion of [Mg/Fe], it is currently
difficult to determine which IMFs 
($\alpha=2.35$ or 2.55 or 2.75) can better explain the observed [Mg/Fe]--[Fe/H]
relation. The observed high [Mg/Fe] ($>0.4$) for higher [Fe/H] ($>-0.7$) can
not be explained by any of the standard models in the present study.

\subsubsection{[Ba/Fe]}

Figure 5 shows that [Ba/Fe] starts to slowly increase with time about 2 Gyr
after the commencement of active star formation in the LMC disk.
This slow
[Ba/Fe] increase is due to the ejection of $s-$process elements from
low-mass ($<3M_{\odot}$) AGB stars. The observed 
systematically high [Ba/Fe] ($>0$ for most
stars) at [Fe/H] $<-1.5$ seems to be better reproduced by the models
with steeper IMFs, though the very high [Ba/Fe] ($>0.2$) 
in some stars at such low metallicities
can not be reproduced at all by any of the standard models in the present study.
The reason for the higher [Ba/Fe] at [Fe/H]$<-1$
in steeper IMFs is that the  numerical ratio  of SNe II with masses
of $\sim 8 M_{\odot}$ to those with $\sim 10 M_{\odot}$ (and thus the contribution
of these SNe to the chemical evolution) 
can change  in the steeper IMF and
consequently [Ba/Fe] can increase more significantly
in the early history  of the chemical evolution of the LMC.
The maximum [Ba/Fe] in the 
model with the Salpeter IMF is at most [Ba/Fe]$\sim 0.4$,
whereas most of the LMC stars at [Fe/H]$>-0.6$
show [Ba/Fe]$ \ge 0.4$.
These results strongly suggest that
the LMC could have had a steeper IMF
(at least steeper than the Salpeter IMF with
$\alpha=2.35$)  in its star formation history
(if the LMC has not experienced starbursts).
The observed very high [Ba/Fe] ($>0.8$) of some stars 
at [Fe/H]$\sim -0.6$ can not be explained at all by any
of the standard models in the present study.

\subsubsection{$f_{\rm g}$-[Fe/H] relation}

In the present study,  the $C_{\rm q}$ for  each model is chosen such that
the final [Fe/H] 
(i.e., [Fe/H]$_0$) is consistent with the observed one ($-0.3$).
A smaller amount of metals can be ejected from SNe in 
the models with steeper IMFs owing to the smaller numbers of SNe.
Therefore, ISM in the models
with steeper IMFs needs to be more rapidly consumed by star formation
and chemically enriched by SN ejecta
to reach [Fe/H]$_0=-0.3$. 
Figure 6 shows that (i)
the models with steeper IMFs show smaller $f_{\rm g,0}$
and (ii) the model with $\alpha=2.55$ is the most
consistent with the observed $f_{\rm g,0}$.  These results suggest that
the observed $f_{\rm g,0}$ and [Fe/H]$_0$ combine to support
the steeper IMF ($\alpha \approx 2.55$) of the LMC.
However this suggestion depends on the adopted assumption that
no ISM can be expelled from the LMC in the standard models.  
If a significant fraction of cold ISM can be stripped from the LMC,
even the models with shallower IMFs (e.g., $\alpha =2.15$) may 
be able to explain both $f_{\rm g,0}$ and [Fe/H]$_0$.
Likewise, if a significant fraction of cold ISM can be recently
accreted onto the LMC,
the models with rather steep IMFs (e.g., $\alpha  >2.75$) 
may explain both $f_{\rm g,0}$ and [Fe/H]$_0$ too.

\subsection{Burst models}

\subsubsection{AMR}

The burst of star formation and the subsequent efficient production of metals 
can rapidly and significantly increase [Fe/H] in the burst models. 
Therefore, the chemical evolution in the quiescent phase of star formation
in the burst models needs to proceed more slowly 
(i.e., smaller $C_{\rm q}$) in comparison with
the standard models so that the final [Fe/H] can be as low as $-0.3$.
Figure 7 shows the AMRs in the five burst models with different IMFs
and $C_{\rm q}$.  The burst model B3 can not reproduce
the observed AMR owing to the systematically low [Fe/H]
in the quiescent phase of star formation.
The model B1 with the Salpeter  IMF can better explain  the AMR of  C05 and C08
whereas the models with steeper IMFs (B2 and B5) can explain
better the AMR by HZ09.
The observed AMR thus cannot give strong constraints on the IMF
of the LMC.
Owing to the observed large [Fe/H] dispersion at 
an age of $\sim 2$ Gyr ($t=11$ Gyr), 
the AMR alone can not allow us to make a robust conclusion
as to whether the LMC experienced a burst of star formation
at an age of 2 Gyr (i.e., $t=11$ Gyr).

\subsubsection{[Mg/Fe]}

As shown in Figure 8,
the overall trends of the [Mg/Fe] evolution in the quiescent phase
for the burst models are similar to those for the standard ones.
The [Mg/Fe]--[Fe/H] relations in the burst models show ``bumps''
(a sharp increase followed by a decrease in the time
evolution of [Mg/Fe])
whose magnitudes depend on $C_{\rm s1}$ and $\alpha$.
The observed locations of the LMC stars in the [Mg/Fe]-[Fe/H] plane
do not show clearly such a bump. However, if the two stars with
[Mg/Fe]$\sim 0.5$ at [Fe/H]$\sim -1.2$ and $-1.0$ are removed from
Figure 8, then the remaining stars appear to show a bump around
$-0.7 <$[Fe/H]$<-0.6$. Furthermore, the apparent lack of stars with
[Mg/Fe]$>0.3$ at [Fe/H]$ \sim -0.4$ suggests that [Mg/Fe] has been decreasing
since [Fe/H]$=-0.5$. Thus the higher [Mg/Fe] at [Fe/H]$\sim -0.6$ could be
evidence of a starburst around $t=11$ Gyr (i.e., 2 Gyr ago). However, 
the observed [Mg/Fe] dispersion at [Fe/H]$\sim -0.6$ is so large that we can not
make a robust conclusion as to whether a starburst occurred 
in the LMC about $t=11$ Gyr (i.e., 2 Gyr ago). 
Also it should be noted that the smaller number of stars at [Fe/H]$>-0.5$
could be responsible for the apparent lack of stars with higher ($>0.3$) [Mg/Fe].

\subsubsection{[Ba/Fe]}

Figure 9 shows that irrespective of the model parameters,
[Ba/Fe] rapidly decreases soon after the starbursts owing
to the ejection of Fe-rich gas from prompt SNe Ia in the
time evolution of [Ba/Fe]. The temporary
[Ba/Fe] decrease  is subsequently followed by its rapid and sharp increase  
due to chemical pollution by AGB ejecta. 
Owing to the rapid [Ba/Fe] increase,  the burst models can show systematically
higher final [Ba/Fe] in comparison with the standard models with no burst.
Therefore, the observed presence of stars with higher [Ba/Fe] ($>  0.5$)
at [Fe/H]$\approx -0.3$
could be more consistent with the burst models. The burst models with
steeper IMFs ($\alpha=2.75$) show [Ba/Fe]$\sim 0.7$ at [Fe/H]$= -0.3$, 
which suggests that the combination of a steeper IMF and a secondary
starburst could be closely associated with the origin of stars with
[Ba/Fe] as large as 0.7. However, very high [Ba/Fe] ($>0.9$) at
[Fe/H]$\sim -0.6$ can not be explained by the present starburst models.
These stars with very high [Ba/Fe] 
could have been formed directly from AGB ejecta that did not mix well with
the ISM.

\subsubsection{$f_{\rm g}$--[Fe/H] relation}

Figure 10 clearly shows that the burst model with the Salpeter IMF
can not reproduce simultaneously the observed $f_{\rm g,0}$ and
[Fe/H]$_0$ owing to efficient  chemical enrichment. On the other hand,
a larger amount of gas needs to be consumed for the ISM to be chemically
enriched to [Fe/H]$\sim -0.3$ in
the models with steep IMFs ($\alpha=2.75$)
so that $f_{\rm g,0}$ can become significantly smaller 
in the model.  The models with $\alpha = 2.55$
yet different star formation histories in the quiescent  phases
can best reproduce both the observed $f_{\rm g,0}$ and [Fe/H]$_0$. 
These results on the IMF-dependences are essentially the same as those
derived for the standard model.

\subsection{Double-burst  models}

\subsubsection{AMR}

Figure 11 shows 
the dependences of the AMRs on IMFs and star formation histories
in the quiescent phase for the double-burst models.
The derived IMF dependence of AMRs are very similar to
those in the burst ones.  The model DB3 with steep IMF ($\alpha =2.75$) and 
less rapid star formation in the quiescent phase is inconsistent with
any observational results on the AMR. The model DB1 with the Salpeter IMF
can better reproduce  the observed AMRs of  C05 and C08 whereas the model DB2
with moderately steep IMF ($\alpha=2.55$) can be better fit to the AMR
by H09.  Although the burst at $t=12.8$ Gyr  is as strong as that at $t=11$ Gyr,
the signature of the burst in the AMR is less clear in the double-burst models.
The observed AMR shows a large [Fe/H] dispersion
in the youngest stellar population (i.e., [Fe/H] ranging from $\sim -0.6$
to $\sim 0$), and the presence of such relatively metal-poor ([Fe/H]$\sim -0.6$)
stars at the present time is  puzzling.
The recent  accretion of metal-poor gas
and the resultant active star formation (before  the burst at $t=12.8$ Gyr)
could result in  the formation of such metal-poor stars and thus
introduce a larger scatter in [Fe/H] at the young populations.

\subsubsection{[Mg/Fe]}

Figure 12 shows that the time evolution of [Mg/Fe] in the double-burst models
is characterized by two occurrences of sharp [Mg/Fe] increase/decrease
after starbursts, which result in two bumps.  The double-burst models
show two bumps  in the [Mg/Fe]-[Fe/H] plane,
though such bumps are not clearly seen 
in the observations. 
The final [Mg/Fe] is larger than 0 for all double-burst models
owing to the last starburst at $t=12.8$ Gyr, which is a feature that
discriminates between
the burst and double-burst models.
The models (DB3, 4 and 5) with steeper IMFs show smaller final [Mg/Fe] and the final
[Mg/Fe] in the model DB1 with $\alpha=2.35$ is more consistent with
the observed one for the LMC field stars.
The observed [Mg/Fe] of the field stars with 
$-0.5<$[Fe/H]$<-0.3$
appears to increase from $\sim 0$ to $0.2$, which may be regarded
as the second bump due to the starburst  at $t=12.8$ Gyr.
The second bump can be shown more clearly in the models DB3 and DB5 with
steeper IMFs ($\alpha=2.75$).
As noted in \S 3.2,  the observed large [Mg/Fe] dispersion at a given [Fe/H]
makes it difficult for us to confirm the presence (or  absence) 
of the first and second  starbursts in the observed [Mg/Fe]--[Fe/H] relation.

\subsubsection{[Ba/Fe]}

Figure 13 shows that [Ba/Fe] rapidly decreases soon
after the first starburst at $t=11$ Gyr
and then increases until the commencement of
the second starburst at $t=12.8$ Gyr  for the double-burst models
with different IMFs (in the time evolution of [Ba/Fe]).
 The increase of [Ba/Fe] after the second starburst
can not be seen, because the 0.2 Gyr difference between the present and the last
starburst is not long enough for AGB stars to have chemically polluted the LMC ISM.
Owing to the first starburst, [Ba/Fe] can become significantly high
(up to $\sim 0.7$) in the double-burst models. However, the maximum
value across all  models is still significantly lower than those
($>0.9$) observed in some stars with [Fe/H]$= -0.6 \sim -0.5$.
The models with steeper IMFs can show higher final [Ba/Fe] and they
can better explain the clear differences in the locations of field stars
in the [Ba/Fe]-[Fe/H] plane between the LMC and the Galaxy.

\subsubsection{$f_{\rm g}$--[Fe/H] relation}

Figure 14 shows that the best model to explain both $f_{\rm g, 0}$
and [Fe/H]$_0$ simultaneously is the one with $\alpha=2.55$ among the
double-burst models (i.e., DB2 and DB4).  This result combined with those in the standard
and burst models strongly suggests that the IMF of the LMC should be
moderately steeper ($\alpha \approx 2.55$) in its long-term star formation
history.  As demonstrated, the models with steeper IMF can better
explain the presence of the LMC field stars with [Ba/Fe]$>0.5$ and higher
[Ba/Fe] of GCs at low metallicities ([Fe/H]$<-1.5$). It should be
noted here that all ejecta from AGB stars and SNe are retained
in the LMC for these non-wind models.  Thus, it can be
concluded that a steeper IMF ($\alpha \sim 2.55$) can better explain
the star formation and chemical evolution histories of the LMC,
{\it as long as  the LMC has not lost a significant amount of its chemically enriched
ISM through stellar winds.} 

\subsection{Wind  models}
\subsubsection{AMR}

Figure 15 shows that (i) the AMRs are not so different 
between selective wind models (W1$-$W4) with different $\alpha$, $C_{\rm q}$,
and $f_{\rm ej}$ and (ii) chemical evolution
proceeds significantly more slowly in the non-selective wind model (W5)
than in the selective ones  until recently ($t \sim 8$ Gyr)
so that the AMR in the non-selective wind model shows
systematically lower [Fe/H] for a given age.
Although the non-selective wind  model can explain the observed [Fe/H]$_0$,
the total gas mass ejected from the LMC ($M_{\rm ej, t}$) for 13 Gyr
is about 3.4 times larger than
the final stellar mass (i.e., $M_{\rm ej, t} = 9.1 \times 10^9 M_{\odot}$
for $M_{\rm s}=2.7 \times 10^9 {\rm M}_{\odot}$) and thus appears to be
too large.
Both already existing ISM and newly synthesized metals can be efficiently 
removed from the LMC in the non-selective wind models so that the chemical evolution
of the LMC can proceed much more slowly. As a result of this, 
a much larger amount of gas can be removed from the LMC until
the stellar metallicity finally becomes [Fe/H]$\approx 0.3$
in the non-selective wind model (W5).
The AMRs in the selective wind models are very similar to the standard
models:
They are more consistent with the observed one
by C05 and C08 than that by HZ09.
This result implies  that the observed AMR alone does not enable us to discuss
the effects of stellar winds in the LMC chemical evolution.

\subsubsection{[Mg/Fe]}

Figure 16 shows that the monolithic decrease of [Mg/Fe] with time
and the [Mg/Fe]--[Fe/H] relation in the five wind models are essentially
similar to those derived in the standard models with no wind. 
The selective wind models with larger $f_{\rm ej}$ show slightly
higher [Mg/Fe] for a given IMF whereas those with steeper IMFs
show lower [Mg/Fe] for a given $f_{\rm ej}$.
The five wind models can not explain the observed higher [Mg/Fe] ($>0.2$)
for [Fe/H] $>-0.6$ in the LMC field stars.  There is no remarkable
difference in the [Mg/Fe] evolution between the selective and non-selective wind models.

\subsubsection{[Ba/Fe]}

Figure 17 shows that the selective wind models can show higher [Ba/Fe] ($>0.5$)
for [Fe/H]$>-0.6$ for a range of $\alpha$ and $f_{\rm ej}$. 
The model W2 with
$\alpha=2.55$ and $f_{\rm ej}=0.4$ shows [Ba/Fe]$>0.6$ without secondary
starbursts,  because Fe-rich gas is ejected from the LMC disk whereas 
the AGB ejecta containing $s$-process elements (e.g., Ba) can be retained and
thus used for chemical evolution. These results imply that efficient
removal of SNe ejecta from the LMC disk is a way to significantly increase
[Ba/Fe] without starburst(s). The models with steeper IMFs
show higher [Ba/Fe], which is essentially the same as
the results derived for  the standard models.
The final (thus maximum) 
[Ba/Fe] in the non-selective wind model at [Fe/H] $=-0.3$ is
lower than 0.4, and therefore  inconsistent with  the observed value,
which implies that the non-selective wind model is less promising
than the selective wind one
as a reasonable model for LMC chemical evolution.

\subsubsection{$f_{\rm g}$--[Fe/H] relation}

Figure 18 shows that the selective wind model W1  with
the Salpeter IMF appears to better fit the observed $f_{\rm g,0}$
and [Fe/H]$_0$ than the models with steeper IMF ($\alpha =2.55$)
for $f_{\rm ej}=0.4$ (W2). However, the model with steeper IMF and $f_{\rm ej}=0.2$
(W4) can also better explain the observed two quantities than that with
the Salpeter IMF and $f_{\rm ej}=0.2$ (W3): it should be noted
that the model W5 can also explain $f_{\rm g,0}$ but does not show the observed high [Ba/Fe]
at [Fe/H]$\sim -0.6$.  These results mean that
if we carefully choose the two parameters ($\alpha$ and $f_{\rm ej}$),
the observation can be well reproduced. The non-selective wind model
with the Salpeter IMF can also reproduce both $f_{\rm g,0}$ and [Fe/H]$_0$ 
reasonably well. 
We do not intend to discuss the wind models with
starburst(s), because the effects of starbursts on the LMC chemical evolution
are essentially the same as those already described in the burst and 
double-burst models. 

\subsection{Comparison between the four different types of models}

We here briefly summarize the advantages and disadvantages of the four
different types of models in reproducing recent observational results
(see   Table 2 for a brief summary).
The observed AMR  can be consistent with most models with
a reasonable set of model parameters.
The models with steeper IMFs
($\alpha \ge 2.55$) can better explain the observed $f_{\rm g,0}$--[Fe/H] relation
and [Ba/Fe]$>-0.2$ at [Fe/H]$>-1.5$ in a self-consistent manner. 
The apparent bump in the observed [Mg/Fe]--[Fe/H] relation 
at [Fe/H]$\sim -0.6$ can be
better reproduced in models with starbursts.
The higher [Ba/Fe] $\sim 0.5$ at [Fe/H]$\sim -0.3$ can be better explained
by non-starburst models (i.e., standard ones) with steeper IMFs
and starburst ones. 
Very high [Ba/Fe] ($>0.9$) at [Fe/H]$\sim -0.6$ can not be explained
by any of the models in the present study.
The significantly higher ($>0.2$) [Ba/Fe] observed in {\it some} 
old, metal-poor GCs ([Fe/H]$<-1.5$) can not be explained by any of these models, either.

The selective wind models with the Salpeter IMF can better explain observations
than the non-wind models with the Salpeter IMF, though
the selective wind models need to assume
an apparently large $f_{\rm ej}=0.4$ 
and can not accurately reproduce  the [Ba/Fe] of the stars in the metal-poor 
GCs with [Fe/H]$<-1.5$.
This leads us to suggest that a steeper IMF is required for explaining
the different observed chemical properties of the LMC in a self-consistent manner.
The removal of metals from SNe can also significantly
increase the [Ba/Fe] (up to $\sim 0.8$)
after a starburst if a large $f_{\rm ej} >0.5$ is chosen 
in the selective wind models. This implies that the selective removal
of SNe ejecta could be partly responsible for the observed rather high
[Ba/Fe] ($\sim 0.8$) in the LMC field stars.
Thus, there are two possible ways to explain the high [Ba/Fe]: One
is star formation directly from  Ba-rich AGB ejecta and the other is significantly
efficient selective removal of SN ejecta.

Thus, as summarized in Table 2,  the four different types of models
can explain 
most of the observed chemical properties of the LMC reasonably well
(except the unusually high [Ba/Fe]), as long as the model parameters are carefully
chosen.  The observed AMR and  overall trends of the [Mg/Fe]--[Fe/H]
(and equally the [$\alpha$/Fe]--[Fe/H]) and the [Ba/Fe]--[Fe/H] relations
can give less strong constraints on the IMF and the star formation
history of the LMC. Table 3 describes how  the four key chemical properties 
of the LMC constrain the IMF and the presence or absence of
past starbursts.

\section{Discussion}

\subsection{A steeper  IMF?}

The present study has shown that the standard, burst, and double-burst
models with steeper IMFs ($\alpha \sim 2.55$) 
can explain well the observed $f_{\rm g}$ and [Fe/H]$_{0}$
in  a  self-consistent manner. 
Furthermore such non-wind models can  not only better  explain
the observed higher [Ba/Fe] at [Fe/H]$< -1.5$ in the LMC
(due  to the steeper IMFs which raise the $r$-process/Fe ratio),
but also can reproduce well the overall dependence
of [Ba/Fe] on [Fe/H].
Although the selective wind models with the Salpeter IMFs ($\alpha=2.35$)
can explain both $f_{\rm g}$ and [Fe/H]$_0$ in a self-consistent manner
for a larger $f_{\rm ej}$ ($\sim 0.4$),
they can not explain so well
the observed [Ba/Fe] at [Fe/H]$\sim -1.5$.
Therefore, the non-wind models with steeper IMFs seems to be slightly
better models for the adopted stellar yields.

Although it seems to be premature 
to observationally determine whether the IMF of the LMC 
is clearly steeper than the Salpeter one,
a number of previous observations have suggested a steeper IMF
in the LMC.
For example,
Mateo (1988) investigated the IMFs of stars with masses ranging
from $0.9M_{\odot}$ to $10.5M_{\odot}$ in the six clusters
of the LMC and found that the slopes
are typically $\alpha \sim 3.5$.
Hill et al. (1994) also found a steeper IMF ($\alpha \sim 3$) for young
stars with masses larger than $9M_{\odot}$ in the LMC and also suggested 
that the IMF can be different below and above $9M_{\odot}$. 
Holtzman et al. (1997)  investigated the IMF for stars on the main sequence
which are fainter than the oldest turnoff and discussed the possibility
of a steep IMF
with $\alpha=2.75$ for the LMC dominated by young populations.  
As shown in Figures 4 and 6 of this paper, 
 rather steep IMFs (e.g., $\alpha \ge 2.95$) cannot be 
consistent with the observed chemical properties and AMR of the LMC.

The only  difference in the non-wind models with steeper IMFs
and the selective wind ones with the Salpeter ones is the [Fe/H]-dependence
of [Ba/Fe] at lower [Fe/H] ($<-1.5$).  Most of the observed field stars
and GCs with [Fe/H]$<-1.5$ in the LMC show [Ba/Fe]$>-0.2$, which is more 
consistent with the non-wind models with steeper IMFs.
However the total number of data points in the observed [Ba/Fe]--[Fe/H]
diagram is sufficiently small that we can not make a robust conclusion as to  whether
the non-wind models or the selective wind ones are better in explaining
the observed [Ba/Fe] trend with [Fe/H].  
Thus it is undoubtedly worthwhile for future spectroscopic observations
to investigate [Ba/Fe] and other $s$-process elements 
of the metal-poor stellar populations
of the LMC with
[Fe/H]$<-1.5$ 
to obtain better constraints on the IMF and the ejection processes
of SN ejecta in the LMC.

\subsection{Evidence for prompt SNe Ia and jet-induced SNe II?} 

The present study first investigated how a prompt SNe Ia influences
the chemical evolution of the LMC and thereby predicted 
that [$\alpha$/Fe] decreases monotonically (until
a secondary starburst occurs) with 
no remarkable plateau in the [$\alpha$/Fe]--[Fe/H] relation
for low [Fe/H].  
We should  first investigate [Mg/Fe]--[Fe/H] relations, firstly
because this element is most reliably determined by the observations among $\alpha$-elements,
and secondly because its yield is most reliably predicted by nucleosynthesis calculations.
The second best $\alpha$-element for this investigation
is Ca. On the other hand, Ti and Si are not so good since 
Ti is not well predicted by SN II nucleosynthesis, 
and the observed determination of the Si abundance involves a large uncertainty.

The observed [Mg/Fe]--[Fe/H] relation does not
show the predicted  trends so clearly partly because of the large dispersion
in [Mg/Fe] at each [Fe/H]. 
Although the physical reasons for the large [Mg/Fe] dispersion are not so clear,
one of the posssible explanations is described as follows.
If different local regions have different star formation histories
(e.g., due to different local gas densities and gas infall rates)
in the LMC, as observed in recent observations (e.g., HZ09 and R12),
then they can have different [Mg/Fe] owing to different chemical enrichment
histories.
Therefore,
the observed  large scatter can be due to
different evolutions of [Mg/Fe] in different
local regions of the LMC.

The observed [Ca/Fe]--[Fe/H] relation can be also used for discussing whether
there is evidence for the prompt SNe Ia playing a significant role
in the chemical evolution of the LMC.
Figure 19 shows the locations of the LMC stars 
on the [Ca/Fe]-[Fe/H] plane as well as the predicted [Ca/Fe]--[Fe/H]
relations for three different models. 
These three models are chosen because they, in combination, show widely
different [Ca/Fe]--[Fe/H] relations of the LMC stars.
It is clear that [Ca/Fe] decreases as [Fe/H] increases without showing
a clear plateau for [Fe/H]$<-1$  in the LMC stars, which is consistent
with the predictions of the present models with prompt SNe Ia.
In addition, the observed locations of the LMC stars on the [Ti/Fe]-[Fe/H]
plane clearly  show a trend for a wide range of metallicity
similar to [Ca/Fe] (e.g., C12). 
However the [Si/Fe]--[Fe/H] relation
does not show such a clear trend 
owing to a larger [Si/Fe] dispersion
for stars with [Fe/H]$<-1.5$ (e.g., Figure 4 in C12).
These results on the [Ca/Fe]--[Fe/H] and [Ti/Fe]--[Fe/H] relations are  
supporting evidence that prompt SNe Ia have influenced the chemical
enrichment history of the LMC.

Figure 19 also shows that there are significant differences in the
distribution of field stars on the [Ca/Fe]-[Fe/H] plane between
the LMC and the Galaxy. The LMC stars with [Fe/H]$>-1$ have systematically
lower [Ca/Fe] in comparison with their Galactic counterparts with 
similar metallicities. The significantly lower [Ca/Fe] in the LMC
stars can not be easily explained by the models even  with
very steep IMFs with $\alpha = 2.95$, though the models with
steeper IMFs can show lower [Ca/Fe]. 
On the other hand, the observations do not show 
significant differences in the locations of field stars on
the [Mg/Fe]-[Fe/H] plane between the LMC and the Galaxy. Given that
the models with rather steep  IMFs ($ \alpha >2.75$) 
show significantly low [Mg/Fe] ($<-0.2$),
the models with unusually steep IMFs ($\alpha > 2.95$)  
are unable to explain both the observed distributions of the LMC field 
stars on the [Ca/Fe]-[Fe/H] and [Mg/Fe]-[Fe/H] planes. So how can we
explain these observations?

If Ca can be expelled  more effectively from the LMC ISM
during explosions of SNe II in comparison with other $\alpha$-elements,
then [Ca/Fe] becomes significantly lower than [Mg/Fe] as chemical evolution
proceeds. We here suggest that nucleosynthesis of jet-induced SNe II
can be responsible for the origin of the rather low [Ca/Fe] as follows.
Shigeyama et al.  (2010) have  recently investigated hydrodynamical  processes
and nucleosynthesis in jet-induced SNe and derived aspherical distributions
of chemical yields of the SNe. 
They have found that both the [Ca/Fe] and the [Mg/Fe] in the ejecta of a jet-induced SN II
can depend strongly on azimuthal angles $\theta$
(where $\theta=0$ corresponds to the direction of the jet).
They have also found that [Mg/Ca] can be lower in lower $\theta$
 in their A3 model in which the total explosion
energy of an aspherical supernova is $10^{52}$ erg and 
the chemical abundances of O, Mg, Fe, Ca, Cr, Mn, and Zn for each $\theta$
are investigated (see their Figure 1 for the $\theta$ dependences).
If the jet-induced SNe II eject with lower $\theta$ 
(where a larger amount of Ca-rich gas exists)
can be more efficiently expelled from the LMC,
then [Ca/Fe] in the LMC becomes more rapidly  lower
(in comparison with [Mg/Fe]) 
as the chemical enrichment proceeds.
Although it is not clear at this stage whether or not this  more efficient
removal of Ca from the LMC is really possible, we here discuss how much
more efficiently Ca should be removed in order to explain the observed
[Ca/Fe]--[Fe/H] relation.

We have investigated the selective wind  
model (W6) in which Ca is removed more efficiently by a factor of 1.75 from
the LMC in comparison with other elements of SNe II ejecta.
Therefore, $M_{\rm w}$ in the equation (5) is rewritten 
in the model W6 as follows:
\begin{equation}
M_{\rm w}= \left\{
\begin{array}{ll}
1.75f_{\rm ej}M_{\rm ej, sn} & \mbox{for Ca} \\
f_{\rm ej}M_{\rm ej, sn}  & \mbox{for other elements} \\
\end{array}
\right.
\end{equation}
The parameter values of $\alpha$, $C_{\rm sq}$, $f_{\rm ej}$ in
model W6 are the same as those adopted in model W2 (i.e.,
$\alpha=2.55$, $C_{\rm q}=0.015$, and $f_{\rm ej}=0.4$).
Figure 19  demonstrates  that model W6 shows  a steeper [Ca/Fe]
decrease with [Fe/H]
and a rather low final [Ca/Fe] ($\sim -0.3$)  at [Fe/H]$\sim -0.3$. 
Furthermore, it is confirmed that
the [Mg/Fe]--[Fe/H] in the model is also consistent with observations.
Although the results of the model are broadly consistent with observations,
it is not clear why
some of the intermediate-age LMC GCs with [Fe/H]$\sim -0.5$
have higher [Ca/Fe] ($\sim 0$)
than the field stars with similar metallicities.
The possible difference in [Ca/Fe] between the GCs and the field stars 
in the LMC could be related to the differences in the formation processes between
field stars and GCs and thus is beyond the scope of this paper. We will
discuss the origin of this intriguing difference in our forthcoming papers
based on chemodynamical numerical simulations of the LMC.

\subsection{Chemical signatures for the past bursts of star formation}

Recent observational studies have investigated the AMR for different
local regions in the LMC and thereby discussed the spatially resolved
star formation and chemical evolution histories of the LMC 
(e.g., HZ09 and R12). HZ09 have found that there are peaks of star formation
at roughly 2 Gyr, 500 Myr, 100 Myr, and 12 Myr ago in the LMC.
R12 also have revealed the presence of peaks in the star formation
rates at 2.0 Gyr and 250 Myr ago. Recent numerical simulations on
the formation of the Magellanic Stream have shown that the LMC and the
SMC could have experienced strong tidal interactions at about 2 Gyr and 250 Myr ago,
and suggested that the two interactions could have significantly enhanced
star formation in the LMC (DB12).
These recent observational and theoretical studies imply that
the LMC might have experienced a burst of star formation at least twice,
though the epoch and the strength of each burst have not been precisely
determined yet. In the following discussion, we focus on the possible
starbursts at $\sim 2$ Gyr and $\sim 200$ Myr ago.

As shown in previous theoretical models,
past starburst events can be imprinted on the chemical abundances 
of the stellar and gaseous components of the LMC 
(e.g., Russell \& Dopita  1992; T95). In particular, [$\alpha$/Fe] as a function
of [Fe/H] can significantly change during starburst events (e.g., T95) and thus
can be used to give strong constraints on the epochs  of the events.
PT98 clearly showed that (i) the non-selective  wind model with
a secondary starburst about 2 Gyr ago can better explain the observed AMR
and (ii) [$\alpha$/Fe] can significantly and rapidly increase
during the starburst and then slowly decrease to the solar value
(e.g., [Mg/Fe]$\sim 0$). The present study has predicted that
the chemical signatures of the starburst about 2 Gyr ago
include (i) a rapid increase of [$\alpha$/Fe] followed by a  rapid 
decrease, (ii) a rapid decrease of [Ba/Fe] followed by a rapid increase,
and (iii) a time delay between the [$\alpha$/Fe] and [Ba/Fe] peaks.

However, as pointed out in previous sections,   these three
chemical signatures of the past starburst event around 2 Gyr ago
can not be clearly seen in the observational results. 
Figure 20 shows [Mg/Fe]--[Fe/H] and [Ba/Fe]--[Fe/H] relations for the five
burst models (B4$-$B9) with different epochs and strengths of starburst.
The results shown in this figure (and Figure 8) suggest that
although the observed 
higher [Mg/Fe] ($>0.2$) at [Fe/H]$\sim -0.6$ can be consistent
with the rapid increase of [Mg/Fe] due to the starburst event,
the large scatter in [Mg/Fe] around [Fe/H]$\sim -0.6$ does not allow
us to make a robust conclusion on the origin of the higher [Mg/Fe].
Similarly,  the observed larger [Ba/Fe] ($>0.5$) for [Fe/H]$>-0.4$,
which is consistent with the present starburst models,
can not be regarded as strong evidence for the presence of a starburst
about 2 Gyr ago owing to the small number of stars with known
[Ba/Fe] for [Fe/H] $>-0.4$.
More observational data sets are necessary to investigate how  the secondary
starburst about 2 Gyr ago 
might have changed the chemical evolution history of the LMC. 
The AMRs derived by different observational studies are significantly different
so that the comparison between the observed AMRs and the simulated ones
in the present study can not provide strong constraints on the nature
of the past starburst events either.

Concerning a possible starburst about 0.2 Gyr ago,
a chemical signature of the starburst about 0.2 Gyr ago
is the sudden increase of [Mg/Fe] around [Fe/H]$\sim -0.4$, as shown in
the present double-burst models. There is a hint of such a [Mg/Fe] increase
in the observational data for the LMC field stars by P08, 
though the number  of the stars is very small (only three). 
If the starburst is strong enough, then it can be imprinted on the
MDF of the LMC.
Figure 21 describes  the MDFs for  [Fe/H] and [Mg/Fe]
in  the five double-burst models. 
The MDFs here are relative frequency histograms that are binned with 0.1 dex bin width
and normalized to the most populated bin.
Clearly the MDF of [Fe/H]
show the double peaks for the five models and such double peaks
are not observed in the previous observation by C08. 
The observed apparent lack of a bimodal MDF in C08
could be due to the small number of young and metal-rich samples in
C08. If the observational result is real, then it means  that
the starburst about 0.2 Gyr ago is much weaker than modeled in the present
study (so that it can not be detected in the observed MDF).

\subsection{The origin of the dip in the AMR} 

The AMR derived for the LMC by HZ09 (in their Figure 20) 
appears to show a sudden and significant  decrease of [Fe/H] around 
1.5 Gyr ago followed by a rapid increase of [Fe/H], though
HZ09 did not discuss the origin of the possible ``dip'' in the AMR.
The AMRs for some local regions of the LMC derived by R12 also appear
to show the  dips whereas the AMR by C08 does not clearly show the dip. 
If the observed dip around 1.5 Gyr ago (HZ09) is real, 
it has profound implications on the gas accretion history of the LMC.
One of likely explanations for the possible dip is that
a large amount of external metal-poor
gas ([Fe/H] $<-1.0$, significantly smaller than the gaseous metallicity
of the LMC about 2 Gyr ago)
was accreted onto the LMC from outside the LMC disk
and consequently [Fe/H] rapidly and significantly decreased. 
The observed apparently rapid decrease of [Fe/H] by almost 0.2 dex at
[Fe/H]$\sim -0.7$ (HZ09) can give some constraints on the amount of
gas accreted onto the LMC for a given metallicity of the metal-poor gas.

In order to discuss how the AMR of the LMC can change owing to
the infall of metal-poor gas from outside the LMC disk,
we have investigated the AMRs of models with gas infall 
yet without any starburst before infall.
The purpose of this investigation is to show clearly how the dip of the AMR
can be formed during the infall of metal-poor gas
(thus not to reproduce the observed AMR fully self-consistently).
Therefore, we think that
the adopted somewhat idealized models would be enough to show clearly 
the formation of the dip in the AMR due to the infall of metal-poor gas.
We have  mainly investigated how much  gas needs to be
accreted onto the LMC by using the results of the ``infall models''
in which metal-poor gas with [Fe/H]=$-1.6$
can be accreted onto the LMC at 1.5 Gyr ago. 
Here the metallicity of [Fe/H]$=-1.6$ is chosen just for a representative case
to clearly show the formation of the dip in the AMR
(we also investigated the infall models with [Fe/H]$=-1.0$ for comparison).

In the infall models, the following infall rate of external gas
($\dot{M}_{\rm ext}$) is added to the right side of the equation (1):
\begin{equation}
\dot{M}_{\rm ext}= \frac { M_{\rm ext} } {t_{\rm in,e} - t_{\rm in, s} },
\end{equation}
where $M_{\rm ext}$ is the total mass of the external gas 
that can infall onto 
the LMC for $t_{\rm in, s} \le t \le t_{\rm in, e}$, where
$t_{\rm in, s}$ and $t_{\rm in, e}$ are the epochs when the gas infall
starts and ends, respectively.  Therefore, the gas infall rate is assumed to
be steady and constant in the present infall models. We investigate the models
with $t_{\rm in, s}=11.5$ Gyr and $t_{\rm in, e}=12.0$ Gyr.
The chemical abundance patterns 
of the gas is  
assumed to be the same as that  of the LMC halo. Therefore, the term
of $Z_{I, i}\dot{M}_{\rm ext}$ 
(where $Z_{I, i}$ is the chemical abundance of each element in the infalling
gas)
is added to the right hand side
of equation (2) to calculate the evolution of $Z_{i}$.
Owing to the infall of metal-poor gas,   the mean metallicity of the youngest
population becomes significantly lower. Therefore, a starburst 
needs to occur after the gas infall 
so that the final [Fe/H] can be $-0.3$, as observed.
We thus assume that starbursts can occur before and after 
accretion of the metal-poor gas onto the LMC disk in  
the  infall models, because the models with this assumption
can be consistent also with the observational results by HZ09
(i.e., a possible starburst about 2 Gyr ago).
We have particularly investigated the  four infall models (I1$-$I4)
with $\alpha=2.55$
and $C_{\rm q}=0.004$, $C_{\rm sb1}=0.8$, $t_{\rm sb1, s}=11.0$ Gyr,
$t_{\rm sb1, e}=11.1$ Gyr,
$t_{\rm sb2, s}=12.0$, $t_{\rm sb2, e}=13.0$ Gyr, 
[Fe/H]$=-1.6$ in external infalling gas, and
different $C_{\rm sb2}$ and $M_{\rm ext}$.
These models were chosen because they can together show different degrees
of sudden [Fe/H] drop (or different depths of the dip) in the AMRs.
For comparison, we have investigated a model (I5) with [Fe/H]$=-1.0$
in external gas for comparison. The model parameters for these models
are shown in Table 4.

Figure 22 shows the AMRs for  the last $\sim 5$
Gyr (i.e., $t=8-13$ Gyr) in the four infall models in which 
$M_{\rm ext}$ ranges from 0.1
to 0.4 in simulation units. 
$M_{\rm ext}=1$ thus means that the total amount of external gas accretion
is the same as the total amount of gas accreted onto the  LMC
from  its own halo for the last 13 Gyr.
The value of $C_{\rm sb2}$ in each model is chosen such that
the final [Fe/H]  can be consistent with the 
observed one.
For comparison, the AMR of
the burst model B2 without gas infall is shown in this figure.
Clearly stellar [Fe/H] can rapidly decrease soon after the metal-poor
gas is accreted onto the LMC with the magnitude of the decrease
depending on the amount of the accreted gas ($M_{\rm ext}$).
The  model I3  with $M_{\rm ext}=0.3$ 
can show 
the dip of $\sim 0.2$ dex, which means that 
$\sim 10^9 M_{\odot}$ needs to be accreted onto the LMC for explaining
the observed dip.
Other  models  with lower $M_{\rm ext}$ (i.e.,  I1 and I2)
show less remarkable dips in the AMRs and thus are less consistent
with the observations. 
If the metallicity of the accreted gas is higher than the adopted
one ([Fe/H]=$-1.6$), then even a larger amount of gas needs to be 
accreted onto the LMC to form the remarkable dip in the AMR.
For example, if the infalling gas has [Fe/H]$=-1$, then
$M_{\rm ext} \approx 0.4$ is required for explaining
the dip of $\sim 0.2$ dex observed in the AMR.
We thus conclude that if the observed dip is real,
then a massive accretion
event of metal-poor gas with $M_{\rm ext}$ of 
at least $\sim 10^9 M_{\odot}$ is required to explain the 
observed magnitude of the dip.

Previous numerical simulations by BC07 and
DB12 demonstrated that 
the gas of the SMC can be accreted onto the LMC after 
tidal stripping of the SMC gas caused by the strong LMC--SMC--Galaxy
interaction. 
These simulations showed that there could be two accretion events 
around 1.5 Gyr and 0.2 Gyr ago, which means that the first accretion
event about 1.5 Gyr ago is a promising candidate 
which can provide a large enough amount of gas
to form  the dip in the LMC AMR.   However, the total amount of the SMC
gas transferred to the LMC about 1.5 Gyr ago in
these simulations is less than 
$10^8 {\rm M}_{\odot}$ for the total SMC mass of
$3 \times 10^9 M_{\odot}$. The predicted gas mass is much smaller
than the required mass ($M_{\rm ext} \sim 10^9 M_{\odot}$) for 
explaining the observed dip. 
This means that
the SMC gas accreted onto the LMC is unlikely to explain the observed dip of 0.2 dex,
if the total mass of the SMC about 2 Gyr ago is similar to
that of the present SMC ($\sim 3 \times 10^9 M_{\odot}$).
However,  if the SMC was
originally much more massive,  the required amount of gas mass could
be transferred from the SMC to the LMC.

Then where does such a large amount of metal-poor gas come from?
One possible scenario is that a gas-rich dwarf galaxy with a total
gas mass of $10^9 M_{\odot}$ merged with
the LMC about 1.5 Gyr ago and the mixing of the gas with the LMC ISM
caused a significant decrease of [Fe/H]. Since the gas-mass of the dwarf
should be similar to that of the LMC in order to explain 
the observed dip,  the dwarf would have
to have had a total mass similar to the LMC.  This means that the LMC stellar disk
could have been severely damaged by the violent dynamical
process of such a major merger event. The thick disk and counter-rotating
stellar components (e.g., Subramaniam \& Prabhu 2005)
might have been formed from this major merger event
occurring in the LMC about 1.5 Gyr ago.
A merged dwarf as massive as the LMC should have a much lower
star formation rate than the LMC so that it can have a low metallicity.
This discussion depends on
the assumption that the observed dip of 0.2 dex is real and the dip
was caused by a massive gas accretion event. 
Given that the amplitude of the dip provides information about
the total mass of a gas-rich dwarf merging with the LMC
(or the total mass of gas accretion onto the LMC), it would be 
quite important for
future extensive observational studies to confirm
the presence or the absence of the dip(s) in the AMR.

\subsection{Formation of very high [Ba/Fe] stars in the LMC} 

Although the present models can reproduce reasonably well
the overall trend of [Ba/Fe] with [Fe/H] and the  higher [Ba/Fe] ($\sim 0.5$)
for [Fe/H]$<-0.6$ in the LMC,  they can not explain the stars
with [Ba/Fe]$>0.9$. We have confirmed that
even the selective wind models with steep IMFs 
and efficient metal ejection ($f_{\rm ej}\sim 0.6$) can show 
at most [Ba/Fe]$\sim 0.8$. 
 This failure of the present one-zone chemical
evolution models is related to the adopted assumption that AGB ejecta
can be mixed with ISM soon after AGB stars eject their $s$-process elements.
If the AGB ejecta does not mix well with the surrounding ISM and consequently 
can be converted into new stars, then the stars can show rather high [Ba/Fe]:
Figure 1 in TB12 shows the observed  rather high ($>1$) 
[Ba/Fe] in the envelope  of AGB stars with different metallicities.
We thus propose  that the stars with unusually high [Ba/Fe] ($>0.9$)  in the LMC
were formed as a result of incomplete mixing of ISM and AGB ejecta.
A key question here is how energetic stellar winds from AGB stars can cool
down to become cold gas for star formation
without mixing so well with the surrounding ISM.

Recent hydrodynamical
simulations have shown that secondary star formation directly from AGB ejecta is
possible in massive star clusters owing to the deeper gravitational
potentials  (e.g., Bekki 2011). 
Although this secondary star formation appears to be
a convincing mechanism  for the formation
of stars with unusually high [Ba/Fe] in the LMC GCs,
it can not explain why  some of
the LMC {\it field stars} show such high [Ba/Fe]. One possibility for
the field star formation from AGB ejecta is that  AGB ejecta can assemble
in the HI holes (e.g., Kim et al. 1999), where ISM can be almost completely
blown away by SNe,  and then can be converted into cold gas there without
mixing efficiently  with chemically enriched ISM. 
The new stars formed from AGB ejecta
in the HI holes can naturally have very high [Ba/Fe]. Thus it would be 
important for observational studies to confirm whether the locations 
of young stars with very high [Ba/Fe] are more likely to be 
in the present HI holes.
Our future chemodynamical simulations will investigate whether this
star formation from AGB ejecta in HI holes is really possible
in the LMC.

\section{Conclusions}

We have investigated the chemical evolution of the LMC by using 
new one-zone chemical evolution models in which both chemical pollution
by prompt SNe Ia 
and metallicity dependent chemical yields of AGB stars
are incorporated and  the IMF is a free parameter.
We have particularly investigated three different types of models
with or without starburst (i.e., standard, burst, and double-burst
models)
so that 
we can discuss the importance of previous starbursts 
in the history of the chemical evolution of
the LMC.
We furthermore have investigated the wind models in which 
gaseous ejecta from AGB stars, SNe Ia, and SNe II
can be removed partly from the LMC owing to
stellar feedback effects of SNe and AGB stars.
The principle results of the models are summarized as follows.

(1) The observed gas mass fraction ($f_{\rm g,0} \sim 0.3$) and
the metallicity of youngest stellar populations
([Fe/H]$_0 \sim -0.3$)  in the LMC
together give some constraints on the IMF and the efficiency of gas removal
by stellar feedback effects.   
Both $f_{\rm g}$ and [Fe/H]$_0$ can be best reproduced by
a  steeper IMF
with $\alpha=2.55$ for the standard (i.e., no burst) models
in a self-consistent manner, if the gaseous ejecta from AGB stars
and SNe are not removed from the LMC (i.e., non-wind models).
This tendency of steeper IMFs to better explain $f_{\rm g}$ and 
[Fe/H]$_0$ simultaneously
can be seen also in the burst and double-burst models.
Furthermore, the observed higher [Ba/Fe] ($>-0.2$)
of stars in GCs
at [Fe/H]$<-1.5$ in the LMC is also more  consistent with 
steeper  IMFs ($\alpha>2.55$).

(2) However, the models with the Salpeter IMF ($\alpha=2.35$) can
also explain the observed $f_{\rm g,0}$ and [Fe/H]$_0$, if significant fractions 
($f_{\rm ej} \sim 0.4$) 
of gaseous ejecta from SNe are {\it selectively} removed from the LMC
(i.e., no removal of AGB ejecta).
These selective wind models are significantly
more reasonable than the uniform wind
ones proposed by PT98 in which
ISM,  AGB ejecta,  and supernova ones are equally removed.
This is because an unreasonably large amount of gas 
($\sim 10^{10} M_{\odot}$, or three times
the present stellar mass of the LMC) need to be removed from the LMC
for the best uniform wind models with the Salpeter IMF.
The wind models with the Salpeter IMF, however, can not reproduce  well
the observed higher [Ba/Fe] ($>-0.2$) at lower [Fe/H] ($<-1.5$).
Thus,  although removal of SNe ejecta could be important in the chemical evolution of the LMC,
an IMF steeper than the Salpeter one
is required for explaining 
$f_{\rm g,0}$, [Fe/H]$_0$, and [Ba/Fe]
at low [Fe/H] in a self-consistent manner.

(3) The present models predict that [$\alpha$/Fe] starts to decrease 
monotonically
only  $\sim 10^8$\ yr after the commencement of active star formation
in the LMC owing to rapid chemical enrichment by ejecta from prompt SNe Ia. 
The observed [Ca/Fe]--[Fe/H] 
relation with the apparent lack of the plateau  can be  consistent with
models with chemical pollution by prompt SNe Ia rather than by classical ones.
This suggests that the observed [Ca/Fe]--[Fe/H] relation
is supporting evidence for prompt SNe Ia playing a key role in  the chemical evolution
of the LMC. However, it should be noted that [Mg/Fe]
does not so show such a clear trend as decreasing [Mg/Fe] with increasing [Fe/H]
(like [Ca/Fe]--[Fe/H] relation)
for the entire sample of stars and GCs.

(4) The present models predict that if the LMC experienced a starburst about
2 Gyr ago,  [$\alpha$/Fe] can start to rapidly increase (up to
0.3) at [Fe/H]$\sim -0.5$ 
owing to gaseous ejecta of SNe II and then soon decrease owing to the Fe-rich
ejecta from prompt SNe Ia. Therefore, the best model predicts a ``bump'' in the
[$\alpha$/Fe]--[Fe/H] relation around [Fe/H]$\sim -0.5$,
though
such a bump can not be so clearly seen in the observed 
[$\alpha$/Fe]--[Fe/H] relation owing to the apparently large scatter
of [$\alpha$/Fe] at [Fe/H]$\sim -0.5$.  
The observed presence of stars 
with [Mg/Fe]$>0.3$ at $-0.6<$[Fe/H]$< -0.5$ and
the observed apparent lack of stars with [Mg/Fe]$<0.2$ at $-0.5<$[Fe/H]$<-0.3$
can be consistent with the presence of the bump,
though the smaller number  of observational data points for [Fe/H]$>-0.5$
could be responsible for the apparent dip.

(5) If the LMC experiences a secondary starburst, [Ba/Fe] temporarily
decreases owing to Fe-rich 
ejecta from SNe II and prompt SNe Ia  then increases sharply owing
to AGB ejecta that is rich in $s$-process elements.
For example,
if the LMC experienced a starburst  about 2 Gyr ago 
when the LMC had [Fe/H]$\sim-0.7$,  then the LMC
shows the [Ba/Fe] peak at [Fe/H]$\sim -0.3$ in
the [Ba/Fe]--[Fe/H] relation. 
Furthermore, [Ba/Fe] can show its peak (up to $\sim 0.7$)
always after [$\alpha$/Fe] takes
its  peak value. As a natural result of this,
[Fe/H] at the [Ba/Fe] peak is higher than that at the [$\alpha$/Fe] peak
in the [Ba/Fe]--[Fe/H] and [$\alpha$/Fe]--[Fe/H] relations. 
These predicted trends of [Ba/Fe] are not so clearly seen in the observed
[Ba/Fe]--[Fe/H] relation owing to the large scatter of [Ba/Fe].
However, the observed larger [Ba/Fe] at [Fe/H] $\sim -0.3$ is more 
consistent with the present burst models (in particular with those of
steeper IMFs).
However, the observed
very large [Ba/Fe] $\sim 1$ at [Fe/H]$\sim -0.5$ 
in some field stars and GCs  of the LMC can not be 
simply explained by any model in the present study.
Thus we have proposed that such very high [Ba/Fe] stars could be 
formed from AGB ejecta that did not mix well with ISM.

(6) The observed AMR has a large scatter,
so the standard (non-burst), burst, and double-burst models
can  be all  consistent with the AMR, which implies that
the AMR does not give strong constraints on the LMC star formation history. 
The present double-burst models have bimodal
distributions of [Fe/H] (i.e., two peaks in the MDF), which have not been observed
yet.  Therefore, the double-burst models are the least consistent
with observations among the three different types of models 
investigated in the present study
in terms of the MDF. Accordingly, we  suggest that the observationally
inferred recent starburst 
around 0.1--0.5 Gyr ago by HZ09 should be weak so as to
reproduce the observed MDF.

(7) If the observed apparent dip 
(i.e., a sudden [Fe/H] decrease by $\sim 0.2$ dex)
in the AMR around 1.5 Gyr ago is real,
then it has a profound implication for the gas accretion history of the LMC. 
The present models predict that the dip could be due to the accretion
of a large amount ($\sim 10^9 M_{\odot}$)
of metal-poor gas ([Fe/H]$<-1$) from other gas-rich galaxies about 1.5 Gyr ago.
Given that previous numerical simulations (e.g., BC07 and DB12) demonstrated
a gas transfer from the SMC to the LMC caused by tidal stripping
of the SMC about 1.5 Gyr ago,
the SMC gas could be responsible for the 
gas accretion event in the LMC. 
However, 
the required large amount of gas ($\sim 10^9 M_{\odot}$) is  much larger
than the predicted amount ($\sim 10^8 M_{\odot}$) in previous simulations.
There can be two possible scenarios for the gas transfer.
One is that
the SMC was originally  much more massive than the present SMC,  as suggested
by the recent modeling of the SMC's rotation curve (Bekki \& Stanivirovic 2009),
so that a large amount of gas can be transferred from the SMC to the LMC.
The other is
that the LMC experienced a major merger with a massive gas-rich dwarf galaxy.
Such a gas-rich major merger event may  be responsible for the formation
of the thicker and extended  stellar disk observed in the LMC.

(8) The observed rather low [Ca/Fe] ($<-0.2$) at [Fe/H]$>-0.6$
 in the LMC field stars can not be
simply explained by the present models, in which 
all elements of  SNe are equally efficiently
removed from the LMC. We have thus proposed that if Ca can be 
(by a factor of $\sim 2$) more efficiently
removed from the LMC through supernova feedback effects
in comparison with other elements, 
then the observed unusually low [Ca/Fe] and normal [Mg/Fe]
for [Fe/H]$>-0.6$  can be simultaneously
explained.  Although it remain unclear why such more efficient removal
of Ca should occur in the chemical enrichment history of the LMC,
we have proposed that a characteristic nucleosynthesis
of jet-induced SNe can be associated with 
the origin of the field stars with unusually low [Ca/Fe].
The origin of the higher [Ca/Fe] in the intermediate-age GCs of the LMC
remains unclear.

(9) The differences in  IMFs and removal efficiencies of AGB and SN ejecta
between the LMC and the Galaxy
might be responsible for 
the observed clear differences in the locations of the stars
in the [$\alpha$/Fe]--[Fe/H] and [Ba/Fe]--[Fe/H] relations between
the two galaxies. Our future chemodynamical simulations of the star formation
and chemical enrichment histories in the LMC will investigate how and why
the IMF and  the removal processes of stellar ejecta
of the two galaxies might be different. We plan to discuss
the spatially different chemical properties 
in the LMC based on the results of the simulations.

\acknowledgments
We are  grateful to the referee  for  constructive and
useful comments that improved this paper.
We are also grateful to Cameron Yozin-Smith for his carefully reading this
manuscript.
KB acknowledges the financial support of the Australian Research Council
throughout the course of this work.

\clearpage

\begin{deluxetable}{ccccccccccc}
\footnotesize  
\tablecaption{Model parameters for one-zone chemical evolution 
\label{tbl-1}}
\tablewidth{-2pt}
\tablehead{
\colhead{  Model  \tablenotemark{a} } &
\colhead{  $\alpha$  \tablenotemark{b} } &
\colhead{  $C_{\rm q}$  \tablenotemark{c} } &
\colhead{  $C_{\rm sb1}$ \tablenotemark{d} }  &
\colhead{  $t_{\rm sb1,s}$  \tablenotemark{e} }  &
\colhead{  $t_{\rm sb1,e}$  \tablenotemark{f} } &
\colhead{  $C_{\rm sb2}$  \tablenotemark{g} }  &
\colhead{  $t_{\rm sb2,s}$ \tablenotemark{h} }  &
\colhead{  $t_{\rm sb2,e}$ \tablenotemark{i}  } &
\colhead{  $f_{\rm ej}$ \tablenotemark{j}   } &
\colhead{  $C_{\rm ej}$ \tablenotemark{k}   }  }
\startdata
S1 & 2.35 & 0.006 & - & - & - & - & - & - & - & -  \\
S2 & 2.15 & 0.004 & - & - & - & - & - & - & - & -  \\
S3 & 2.55 & 0.01 & - & - & - & - & - & - & - & -  \\
S4 & 2.75 & 0.017 & - & - & - & - & - & - & - & -  \\
S5 & 2.95 & 0.02 & - & - & - & - & - & - & - & -  \\
B1 & 2.35 & 0.004 & 0.3 & 11.0 & 11.1 & - & - & - & - & -  \\
B2 & 2.55 & 0.004 & 0.8 & 11.0 & 11.1 & - & - & - & - & -  \\
B3 & 2.75 & 0.004 & 1.2 & 11.0 & 11.1 & - & - & - & - & -  \\
B4 & 2.55 & 0.006 & 0.5 & 11.0 & 11.1 & - & - & - & - & -  \\
B5 & 2.75 & 0.006 & 1.4 & 11.0 & 11.1 & - & - & - & - & -  \\
B6 & 2.35 & 0.004 & 0.03 & 11.0 & 12.0 & - & - & - & - & -  \\
B7 & 2.55 & 0.004 & 0.8 & 10.0 & 10.1 & - & - & - & - & -  \\
B8 & 2.55 & 0.004 & 0.8 & 8.0 & 8.1 & - & - & - & - & -  \\
B9 & 2.35 & 0.004 & 0.6 & 6.0 & 6.1 & - & - & - & - & -  \\
DB1 & 2.35 & 0.004 & 0.2 & 11.0 & 11.1  & 0.2 & 12.8 & 12.9 & - & -  \\
DB2 & 2.55 & 0.004 & 0.55 & 11.0 & 11.1  & 0.55 & 12.8 & 12.9 & - & -  \\
DB3 & 2.75 & 0.004 & 1.0 & 11.0 & 11.1  & 1.0 & 12.8 & 12.9 & - & -  \\
DB4 & 2.55 & 0.006 & 0.4 & 11.0 & 11.1  & 0.4 & 12.8 & 12.9 & - & -  \\
DB5 & 2.75 & 0.006 & 0.9 & 11.0 & 11.1  & 0.9 & 12.8 & 12.9 & - & -  \\
DB6 & 2.35 & 0.004 & 0.15 & 11.0 & 11.1  & 0.15 & 12.5 & 12.6 & - & -  \\
W1 & 2.35 & 0.01 & - & - & - & - & - & - & 0.4 & -  \\
W2 & 2.55 & 0.015 & - & - & - & - & - & - & 0.4 & -  \\
W3 & 2.35 & 0.008 & - & - & - & - & - & - & 0.2 & -  \\
W4 & 2.55 & 0.012 & - & - & - & - & - & - & 0.2 & -  \\
W5 & 2.35 & 0.012 & - & - & - & - & - & - & 0.2 & 300.0  \\
W6 & 2.55 & 0.015 & - & - & - & - & - & - & 0.4 & -  \\
\enddata
\tablenotetext{a}{The ``S'', ``B'', ``DB'', and ``W''  are referred to as
the standard (i.e., no burst), burst, double-burst, and wind models,
respectively. The removal efficiency $f_{\rm ej}$  is by a factor of 1.75 
 higher in Ca than other $\alpha$-elements in the wind model W6.}
\tablenotetext{b}{The slope of the IMF.}
\tablenotetext{c}{The coefficient for star formation in the quiescent
(i.e., no starburst) phase.}
\tablenotetext{d}{The coefficient for star formation in the first 
starburst  phase.}
\tablenotetext{e}{The time at which the first starburst begins in units
of Gyr.}
\tablenotetext{f}{The time at which the first starburst ends in units
of Gyr.}
\tablenotetext{g}{The coefficient for star formation in the second
starburst  phase.}
\tablenotetext{h}{The time at which the second starburst begins in units
of Gyr.}
\end{deluxetable}

\clearpage

\noindent $^{\rm i}$ The time at which the second starburst ends in units
of Gyr. \\
$^{\rm j}$ The mass fraction of SN ejecta that can be completely
removed from the LMC in the selective wind models. \\
$^{\rm k}$ The coefficient for the metal ejection rate in
the non-selective wind models. \\


\begin{deluxetable}{ccccc}
\footnotesize  
\tablecaption{Comparison of different models 
\label{tbl-2}}
\tablewidth{-2pt}
\tablehead{
\colhead{Properties/models \tablenotemark{a} } &
\colhead{Standard}  &
\colhead{Burst}  &
\colhead{Double-burst}  &
\colhead{Wind}  }
\startdata
AMR 
&  $\bigcirc$  & $\bigcirc$  & $\bigcirc$  & $\bigcirc$  \\  
${\rm [Mg/Fe]-[Fe/H]}$
&  $\bigcirc$  & $\bigcirc$  & $\bigcirc$  & $\bigcirc$  \\  
a bump in the  ${\rm [Mg/Fe]-[Fe/H]}$  relation
&  $\times$  & $\bigcirc$  & $\bigcirc$  & $\bigcirc$  \tablenotemark{b}  \\  
${\rm [Mg/Fe]}$ at ${\rm [Fe/H]_0}$  
&  $\bigcirc$  & $\bigcirc$  & $\bigcirc$  & $\bigcirc$  \\  
${\rm [Ba/Fe]-[Fe/H]}$
&  $\bigcirc$  & $\bigcirc$  & $\bigcirc$  & $\bigcirc$  \\  
high ($\sim 0.5$) ${\rm [Ba/Fe]}$ at  ${\rm [Fe/H]} \sim -0.3$ 
&  $\bigcirc$  \tablenotemark{c}  
&  $\bigcirc$  \tablenotemark{c}
&  $\bigcirc$  \tablenotemark{c}
&  $\bigcirc$  \tablenotemark{d} \\
very high ($> 0.9$) ${\rm [Ba/Fe]}$ at  ${\rm [Fe/H]} \sim -0.6$ 
&  $\times$  & $\times$  & $\times$  & $\times$  \\  
${\rm [Ba/Fe]}$($>-0.2$) of GCs at  ${\rm [Fe/H]} < -1.5$ 
&  $\bigcirc$  \tablenotemark{e}
&  $\bigcirc$  \tablenotemark{e}
&  $\bigcirc$  \tablenotemark{e}
&  $\bigcirc$  \tablenotemark{e} \\
$f_{\rm g,0}-{\rm [Fe/H]}_0$ 
&  $\bigcirc$  \tablenotemark{f}
&  $\bigcirc$  \tablenotemark{f}
&  $\bigcirc$  \tablenotemark{f}
&  $\bigcirc$  \\ 
\enddata
\tablenotetext{a}{If a listed chemical property of the LMC
is (not)  reproduced reasonably well
by a model, then a diagnosis mark  ``$\bigcirc$'' (``$\times$'') is given
for the property. }
\tablenotetext{b}{A starburst about 2 Gyr ago is required 
for the (selective and non-selective) wind models}
\tablenotetext{c}{The IMF slope $\alpha$ needs to be steeper than
2.55 in these standard, burst, and double-burst models.}
\tablenotetext{d}{Only selective wind models can reproduce this
observation.}
\tablenotetext{e}{The IMF slope $\alpha$ needs to be steeper than 
2.55 in these four different types of models.}
\tablenotetext{f}{The IMF slope $\alpha$ needs to be  
$\sim 2.55$ in these standard, burst, and double-burst models.}
\end{deluxetable}


\begin{deluxetable}{cc}
\footnotesize  
\tablecaption{Constraints on the IMF and the star formation history
in the LMC
\label{tbl-3}}
\tablewidth{-2pt}
\tablehead{
\colhead{Properties  \tablenotemark{a} } &
\colhead{Requirement}  }
\startdata
a bump in the ${\rm [Mg/Fe]-[Fe/H]}$  relation 
& starburst about 2 Gyr ago \\
high ($\sim 0.5$) ${\rm [Ba/Fe]}$ at  ${\rm [Fe/H]} \sim -0.3$ 
&  $\alpha\ge2.55$ in non-wind models
\tablenotemark{b} \\
${\rm [Ba/Fe]}$($>-0.2$) of GCs at  ${\rm [Fe/H]} < -1.5$ 
&  $\alpha\ge2.55$  \\ 
$f_{\rm g,0}-{\rm [Fe/H]}_0$ 
&  $\alpha\sim 2.55$ in non-wind models  \\
\enddata
\tablenotetext{a}{The listed four properties are selected from
the seven in Table 2 because they can give some constraints on
the IMF and the star formation history of the LMC.}
\tablenotetext{b}{This property can be reproduced in
wind models if only SNe ejecta are partly removed from the LMC
(i.e., only in selective wind models)
for a reasonable range of IMFs.}
\end{deluxetable}


\begin{deluxetable}{cccc}
\footnotesize  
\tablecaption{Parameters for the infall models
\label{tbl-4}}
\tablewidth{-2pt}
\tablehead{
\colhead{Model } &
\colhead{$M_{\rm ext}$  } &
\colhead{[Fe/H] in external gas} &
\colhead{$C_{\rm sb2}$}  }
\startdata
I1 & 0.1 & -1.6 & 0.05    \\  
I2 & 0.2 & -1.6  & 0.07 \\  
I3 & 0.3 & -1.6 & 0.1 \\  
I4 & 0.4 & -1.6 & 0.12 \\  
I5 & 0.4 & -1.0  & 0.12 \\  
\enddata
\end{deluxetable}

\clearpage

\begin{figure}
\plotone{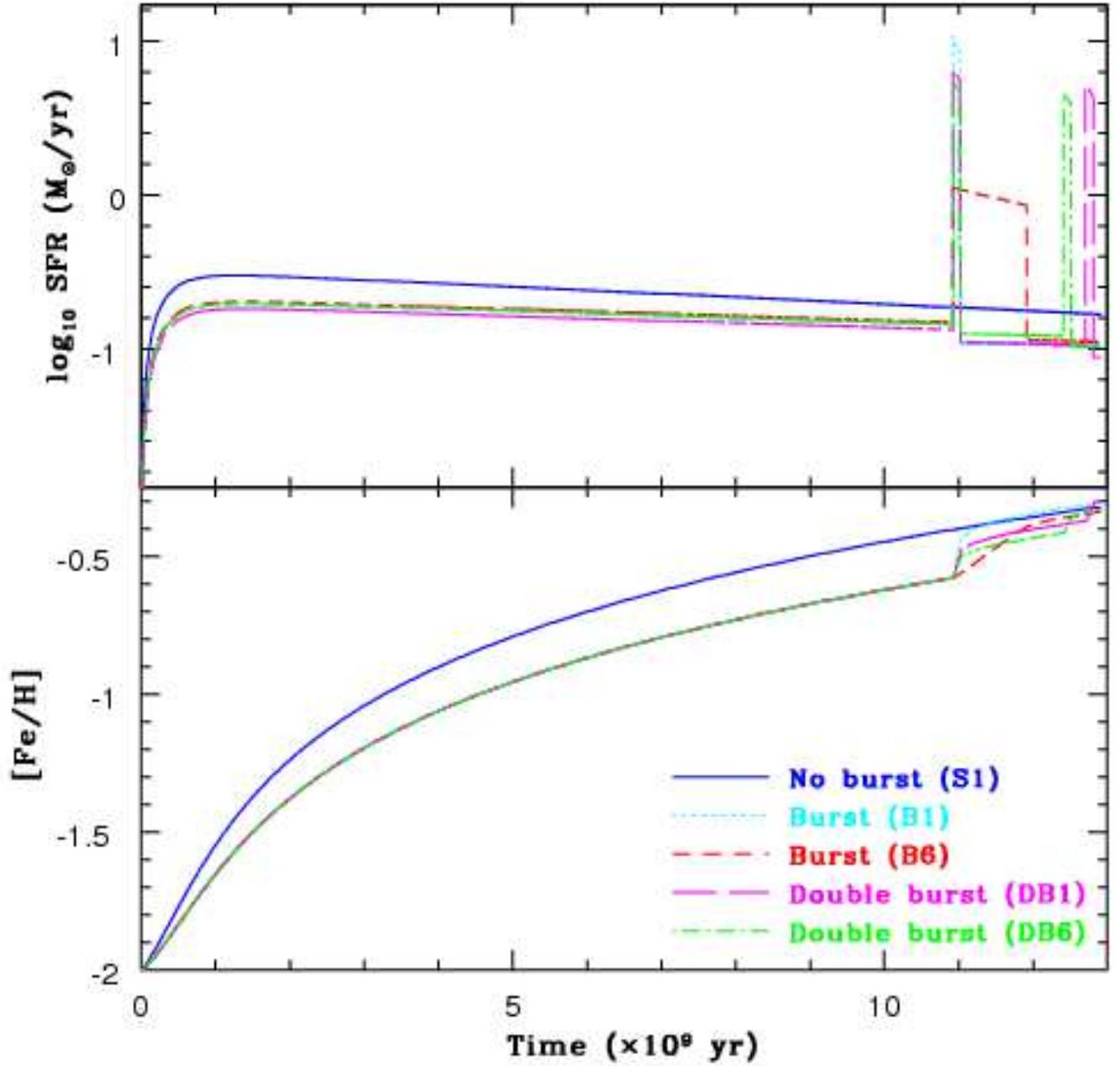}
\figcaption{
The time evolution of star formation rates (SFRs) in
units of $M_{\odot}$ yr$^{-1}$ (upper) and
[Fe/H] (lower) for the five representative models
with different star formation histories, S1 (blue solid),
B1 (cyan dotted), B6 (red short-dashed), DB1 (magenta long-dashed),
and DB 6 (green dot-dashed). The SFRs are estimated for these models
by assuming that
the present total stellar mass of the LMC ($M_{\rm s}$) is 
$2.7 \times 10^9 M_{\odot}$.
\label{fig-1}}
\end{figure}

\begin{figure}
\plotone{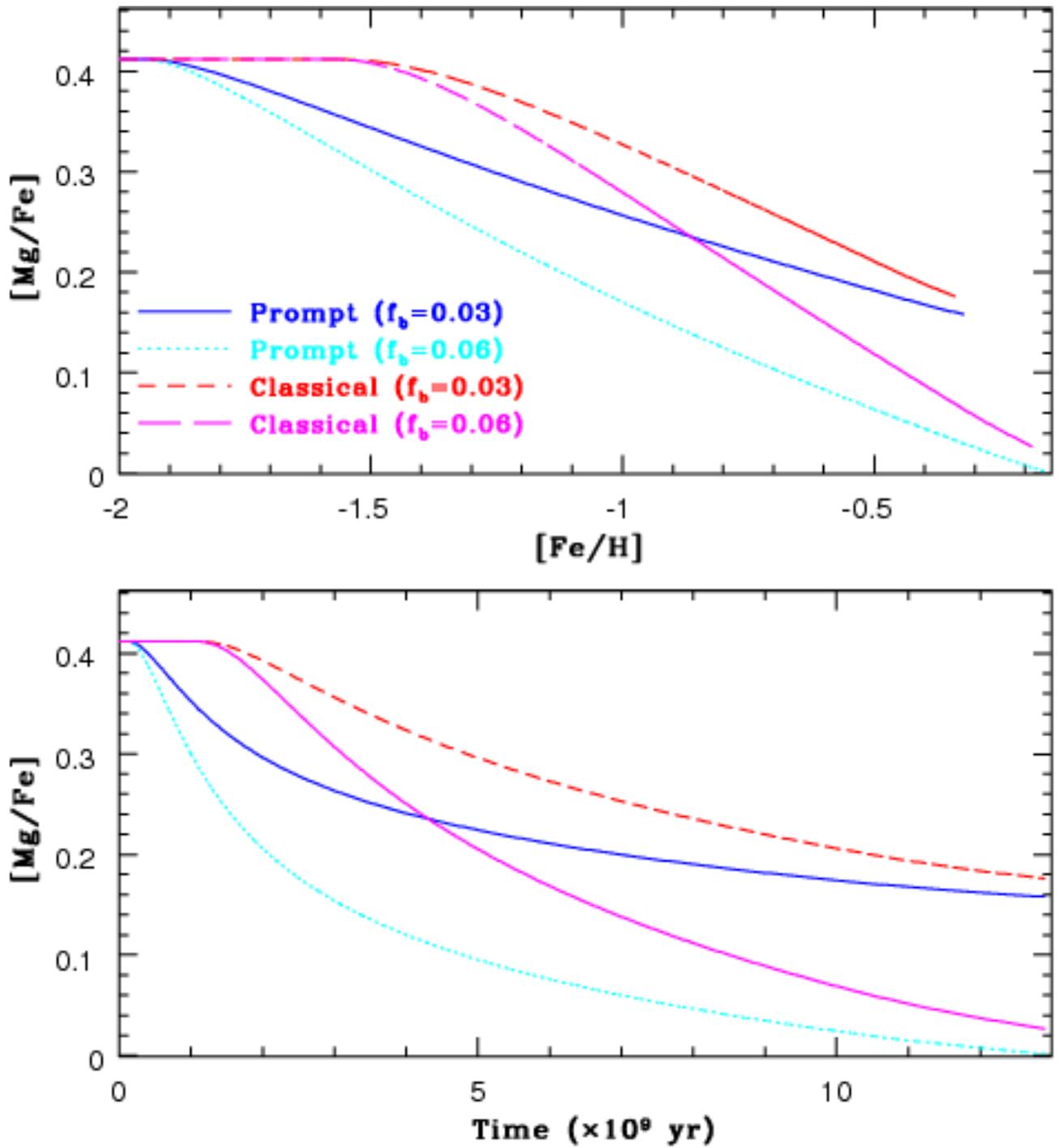}
\figcaption{
Chemical evolution of four models on the [Mg/Fe]-[Fe/H] plane
(upper) and the time evolution of [Mg/Fe] for two prompt SN Ia models
with $f_{\rm b}=0.03$ (blue solid) and 0.06 (cyan dotted)
and two classical SN Ia ones with $f_{\rm b}=0.03$ (red short-dashed)
and 0.06 (magenta long-dashed).
\label{fig-2}}
\end{figure}

\begin{figure}
\plotone{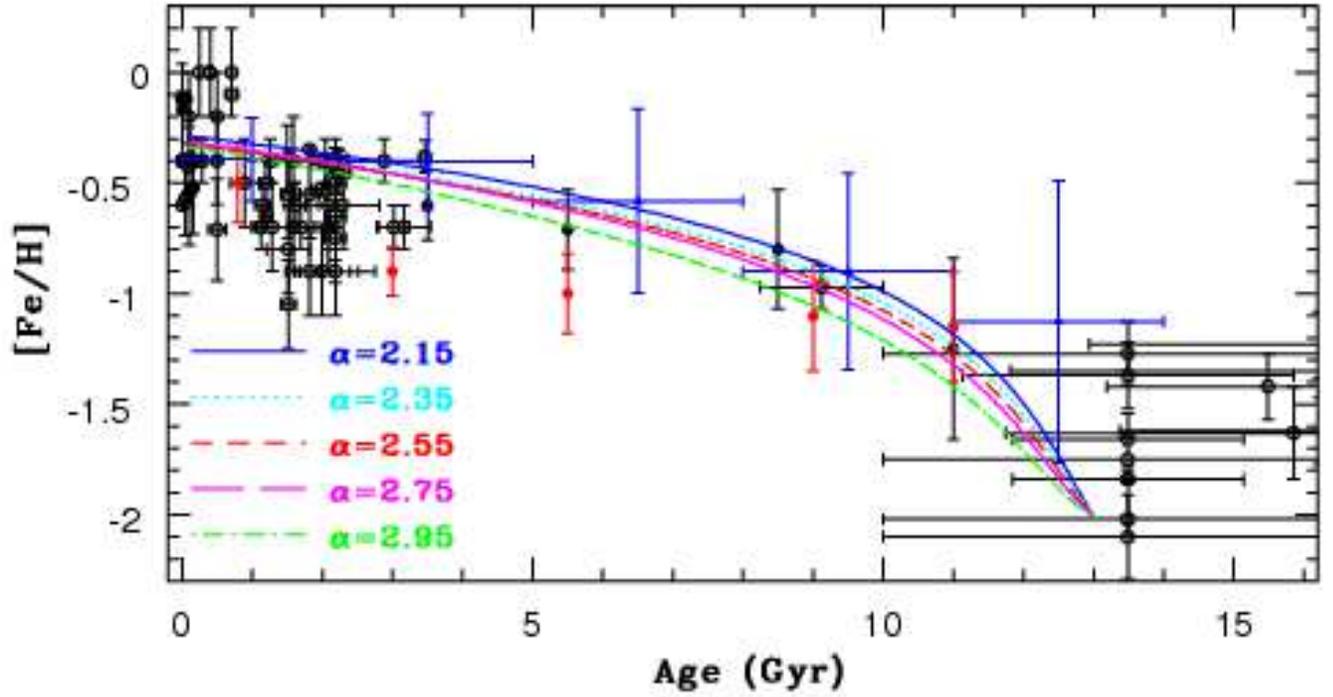}
\figcaption{
The AMRs for the five standard models (S1--S5) with  different $\alpha$
(IMF slope):
$\alpha=2.15$ (blue solid),
$\alpha=2.35$ (cyan dotted),
$\alpha=2.55$ (red short-dashed),
$\alpha=2.75$ (magenta long-dashed),
and $\alpha=2.95$ (green dot-dashed).
Observational results of the LMC field stars
by C05 (blue triangles), C08 (black filled circles),
and HZ09 (red pentagons) are shown.
The AMR for young clusters and GCs from HZ09 are also plotted by
open circles. 
\label{fig-3}}
\end{figure}

\begin{figure}
\plotone{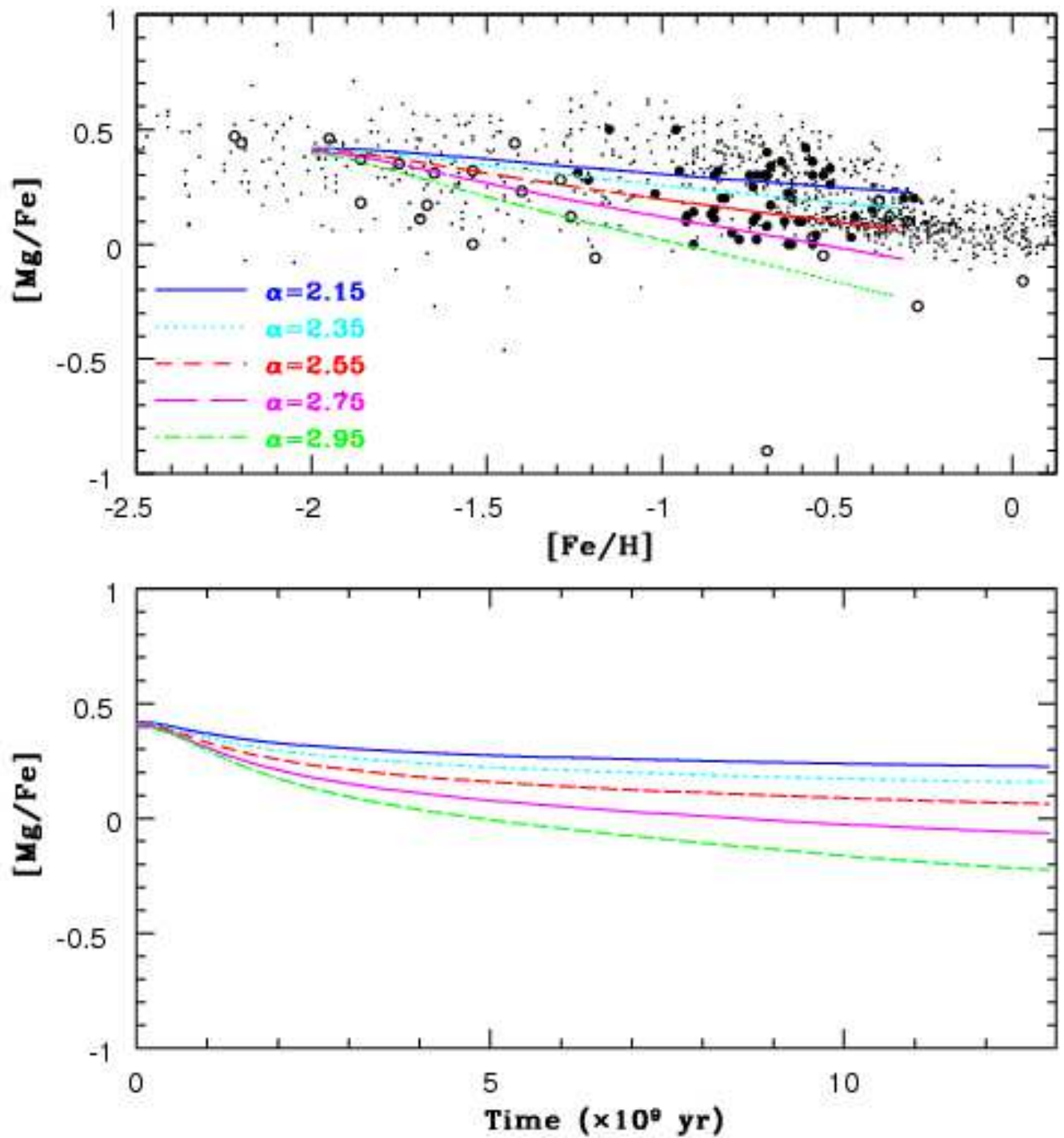}
\figcaption{
Chemical evolution on the LMC disk 
on the [Mg/Fe]-[Fe/H] plane (upper)
and the time evolution of [Mg/Fe] (lower)
for the five standard models (S1$-$S5) with
$\alpha=2.15$ (blue solid),
$\alpha=2.35$ (cyan dotted),
$\alpha=2.55$ (red short-dashed),
$\alpha=2.75$ (magenta long-dashed),
and $\alpha=2.95$ (green dot-dashed).
The observed locations of the  LMC field stars (big filled circles)
and clusters (big open circles) and the Galactic field stars
(small dots) on the [Mg/Fe]-[Fe/H] plane  are shown for comparison.
The observational results  include 
P08
for the LMC field stars,
Johnson et al. 2006 (J06), Mucciarelli et al. (2008, 2010, 2011),  and C12
for the LMC clusters,
Gratton et al. (1999)
Reddy et al. (2003)
Venn et al. (2004), and
Bensby et al. (2005)
for the Galactic field stars.
\label{fig-4}}
\end{figure}

\begin{figure}
\plotone{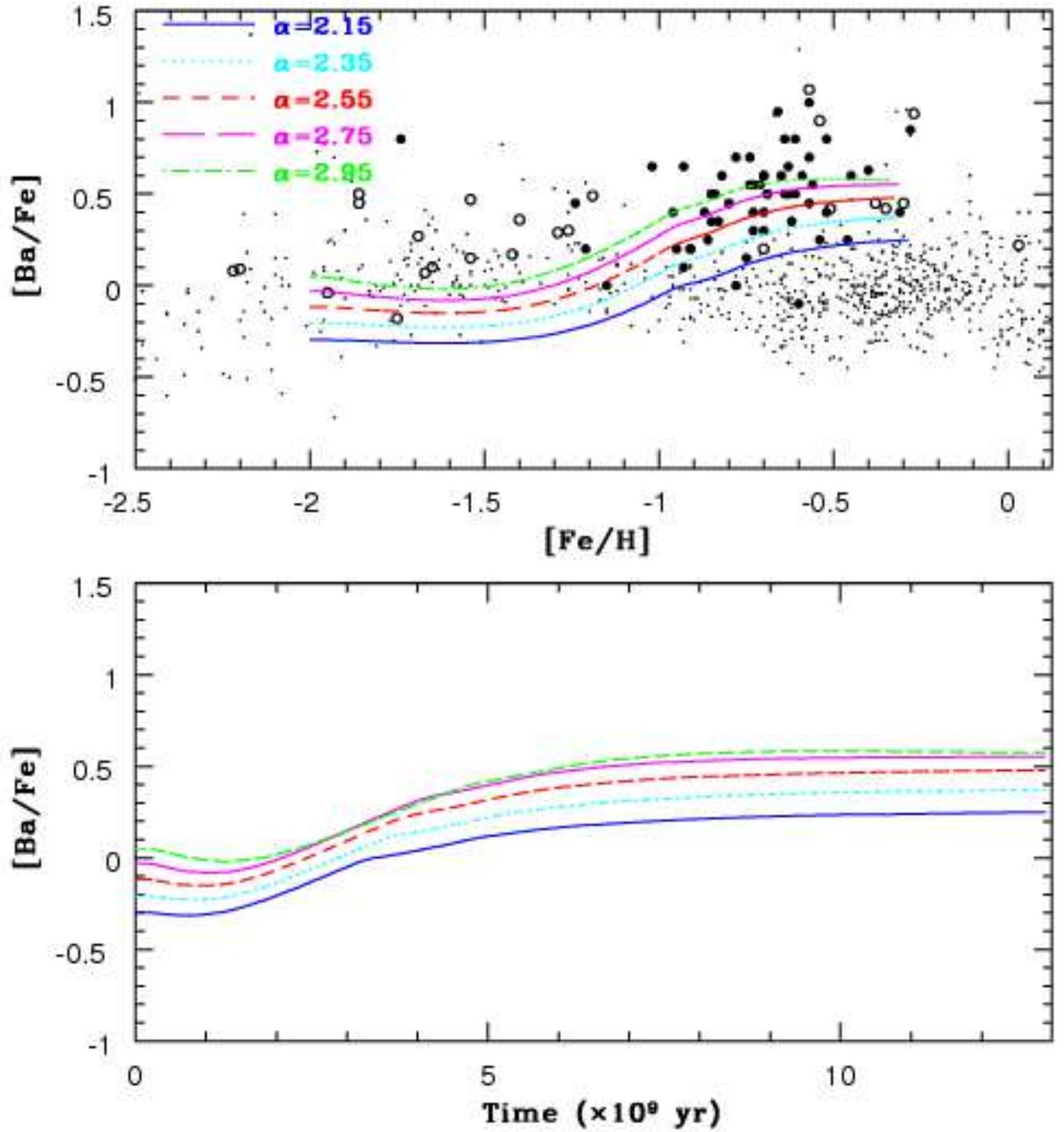}
\figcaption{
The same as Figure 4 but for the [Ba/Fe]--[Fe/H] relations
and the [Ba/Fe] evolution.
\label{fig-5}}
\end{figure}

\begin{figure}
\plotone{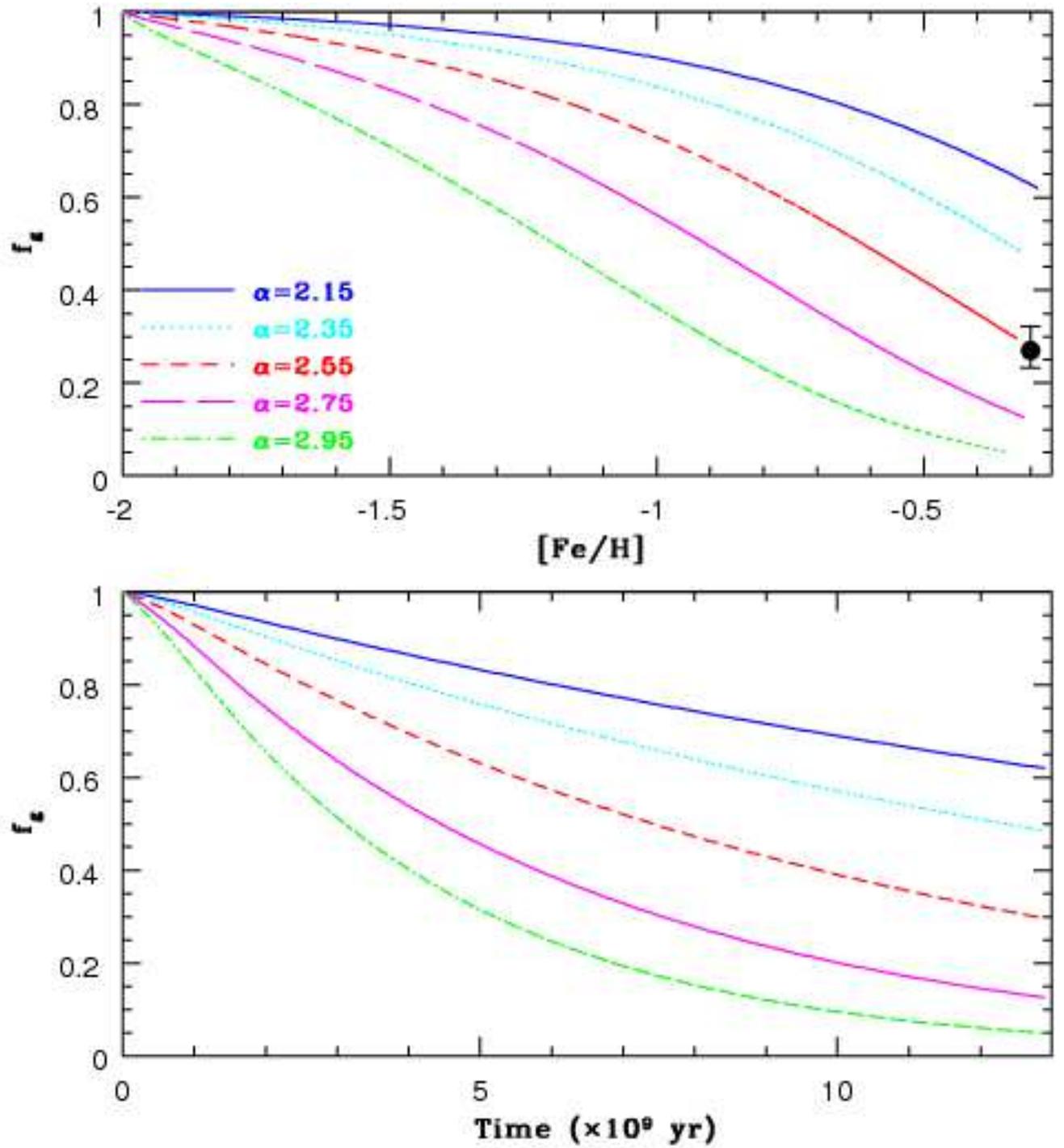}
\figcaption{
The same as Figure 4 but for the $f_{\rm g}$--[Fe/H] relations
and the $f_{\rm g}$  evolution. The filled circle with a vertical
error bar indicates the observed present gas mass fraction
of the LMC ($f_{\rm g, 0}$).
\label{fig-6}}
\end{figure}

\begin{figure}
\plotone{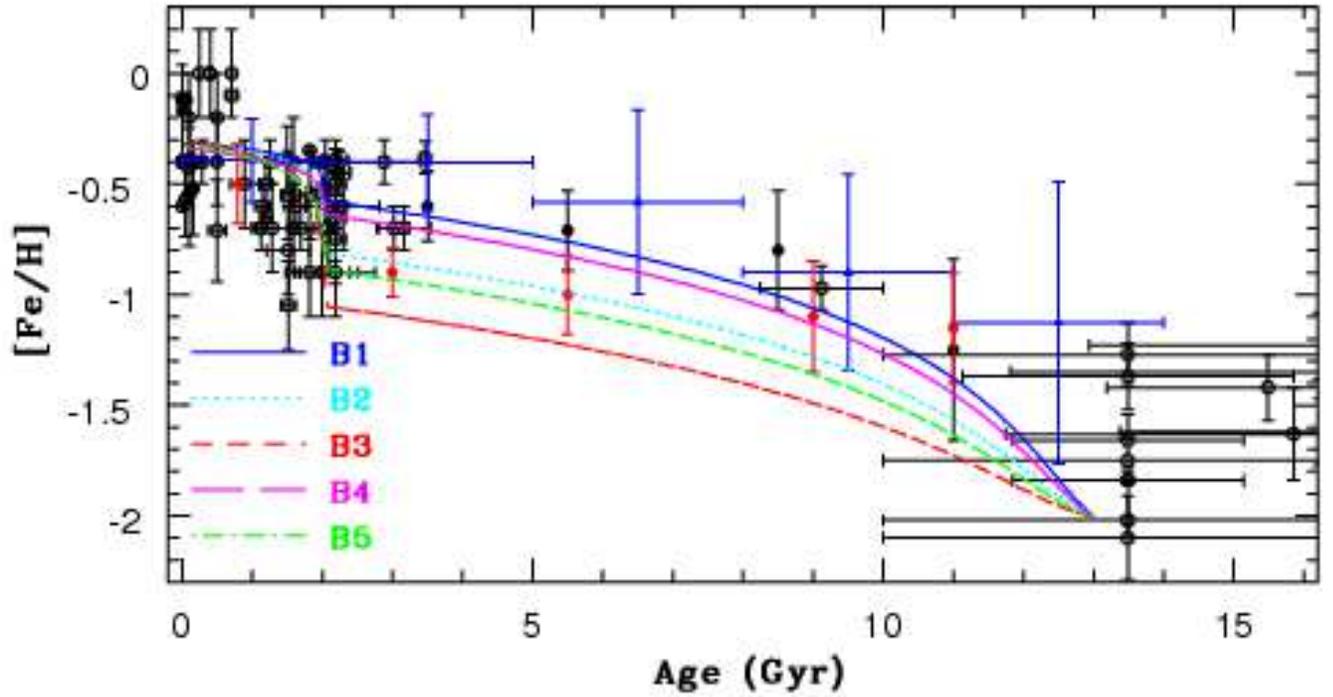}
\figcaption{
The same as Figure 3 but for the five burst models,
B1 with
$\alpha=2.35$ and $C_{\rm q}=0.004$ (blue solid),
B2 with
$\alpha=2.55$ and $C_{\rm q}=0.004$ (cyan dotted),
B3 with
$\alpha=2.75$ and $C_{\rm q}=0.004$ (red short-dashed),
B4 with
$\alpha=2.55$ and $C_{\rm q}=0.006$ (magenta long-dashed),
and B5 with
$\alpha=2.75$ and $C_{\rm q}=0.006$  (green dot-dashed).
The parameter $C_{\rm q}$ can control the rapidity of star formation
in the quiescent phase before starburst (see the main text for more details
of $C_{\rm q}$).
\label{fig-7}}
\end{figure}

\begin{figure}
\plotone{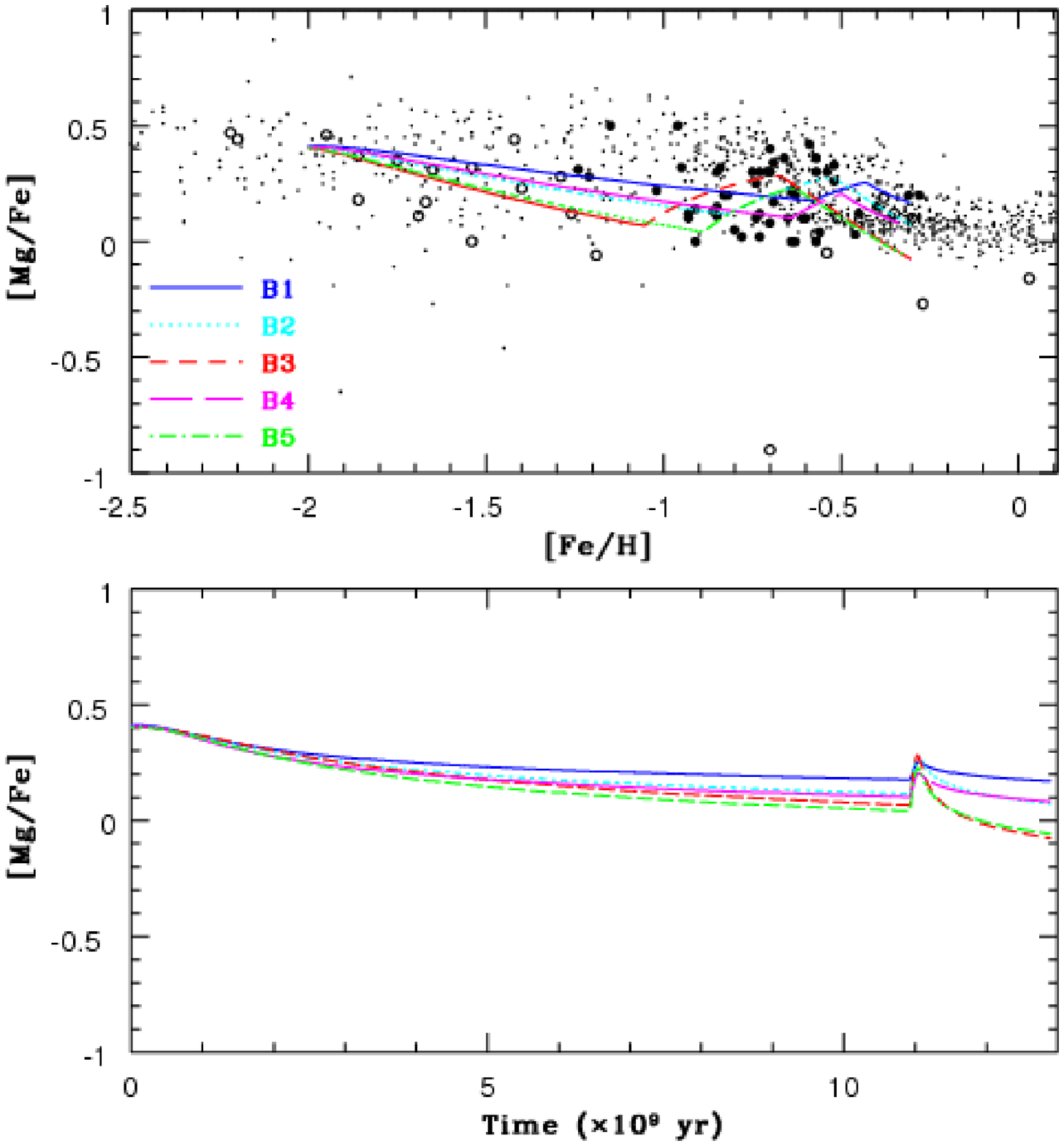}
\figcaption{
The same as Figure 4 but for the five burst models,
B1 with
$\alpha=2.35$ and $C_{\rm q}=0.004$ (blue solid),
B2 with
$\alpha=2.55$ and $C_{\rm q}=0.004$ (cyan dotted),
B3 with
$\alpha=2.75$ and $C_{\rm q}=0.004$(red short-dashed),
B4 with
$\alpha=2.55$ and $C_{\rm q}=0.006$ (magenta long-dashed),
and B5 with
$\alpha=2.75$ and $C_{\rm q}=0.006$  (green dot-dashed).
\label{fig-8}}
\end{figure}

\begin{figure}
\plotone{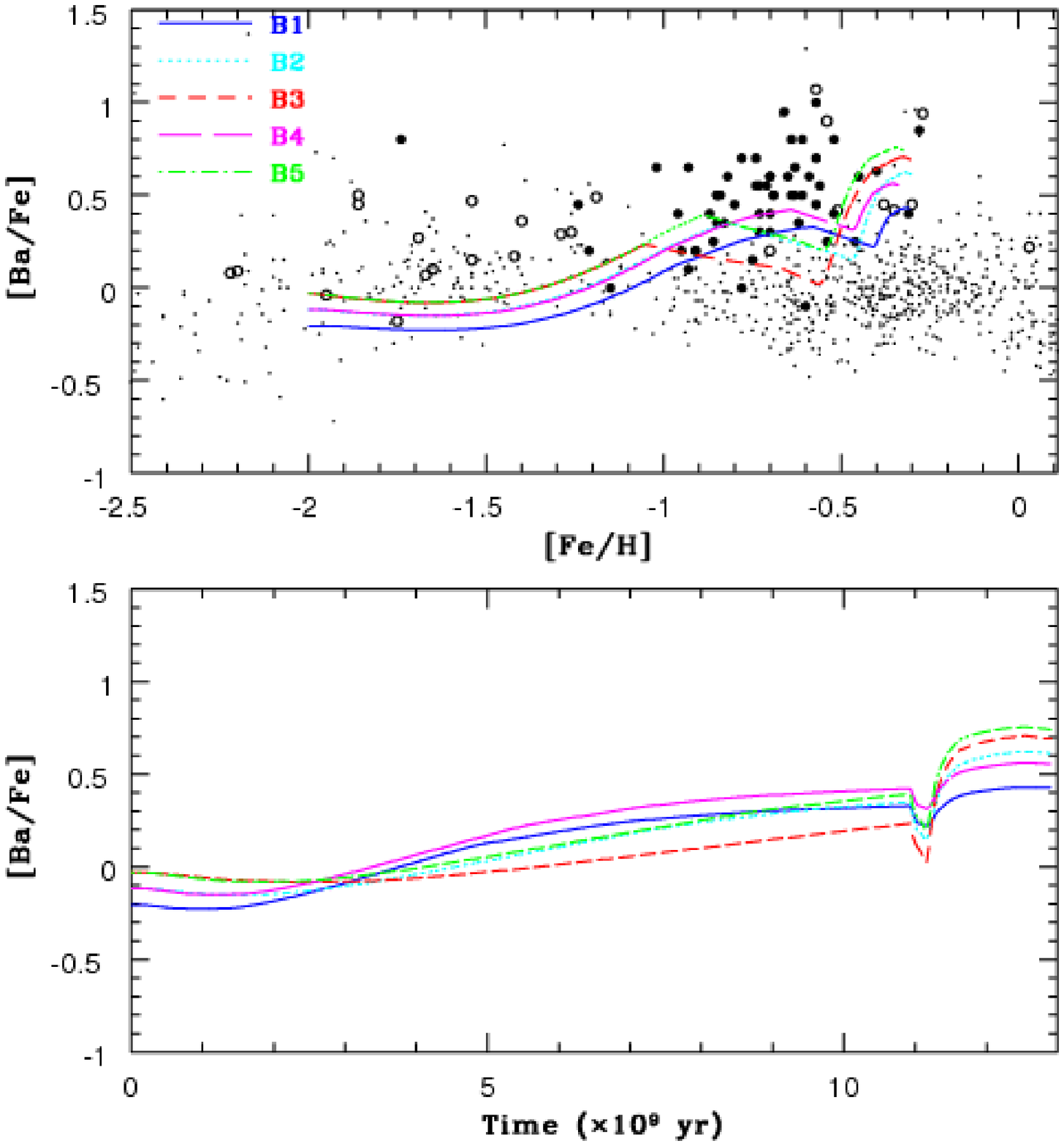}
\figcaption{
The same as Figure 5 but for the five burst models,
B1 with
$\alpha=2.35$ and $C_{\rm q}=0.004$ (blue solid),
B2 with
$\alpha=2.55$ and $C_{\rm q}=0.004$ (cyan dotted),
B3 with
$\alpha=2.75$ and $C_{\rm q}=0.004$(red short-dashed),
B4 with
$\alpha=2.55$ and $C_{\rm q}=0.006$ (magenta long-dashed),
and B5 with
$\alpha=2.75$ and $C_{\rm q}=0.006$  (green dot-dashed).
\label{fig-9}}
\end{figure}

\begin{figure}
\plotone{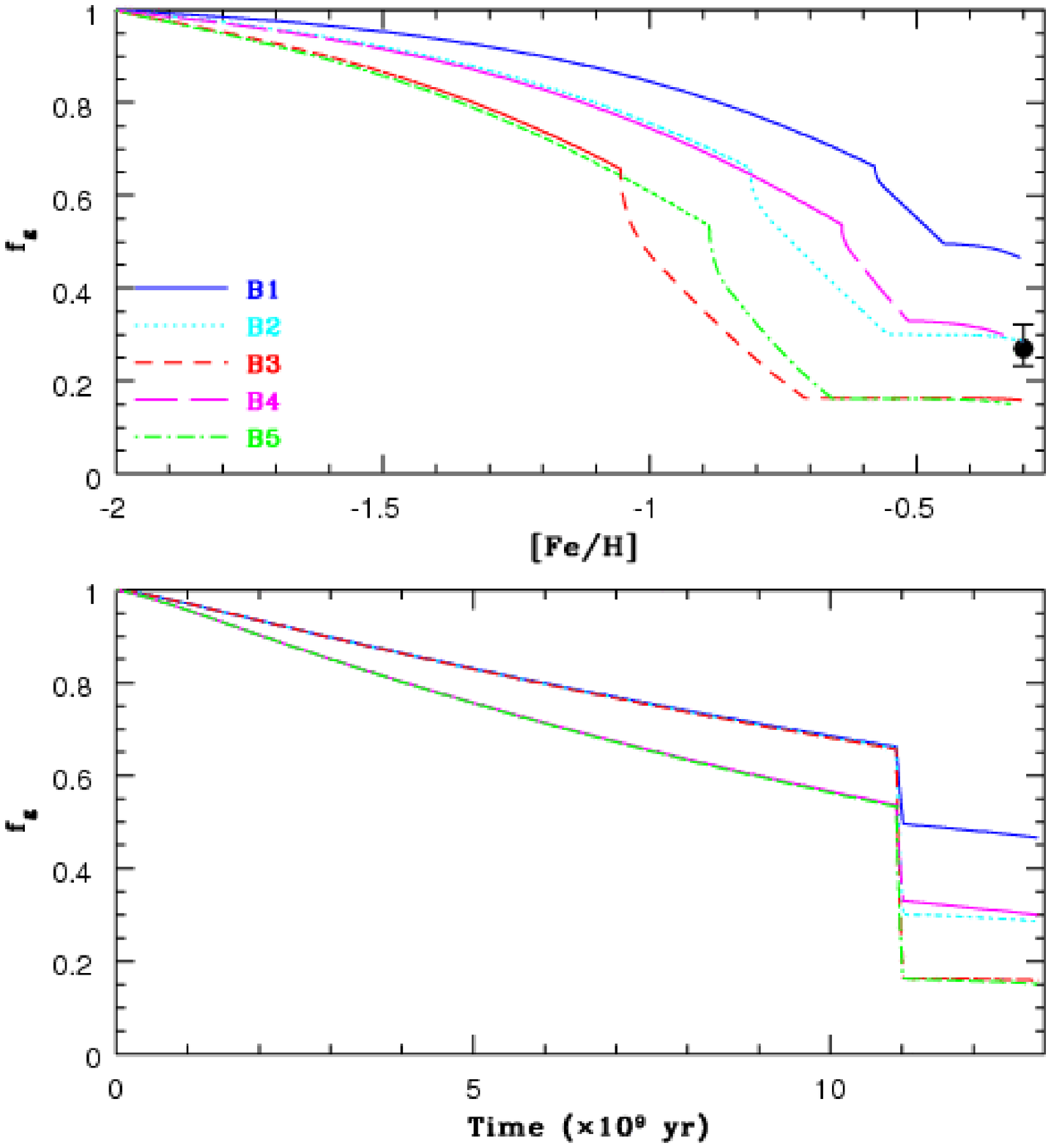}
\figcaption{
The same as Figure 6 but for the five burst models,
B1 with
$\alpha=2.35$ and $C_{\rm q}=0.004$ (blue solid),
B2 with
$\alpha=2.55$ and $C_{\rm q}=0.004$ (cyan dotted),
B3 with
$\alpha=2.75$ and $C_{\rm q}=0.004$(red short-dashed),
B4 with
$\alpha=2.55$ and $C_{\rm q}=0.006$ (magenta long-dashed),
and B5 with
$\alpha=2.75$ and $C_{\rm q}=0.006$  (green dot-dashed).
\label{fig-10}}
\end{figure}

\begin{figure}
\plotone{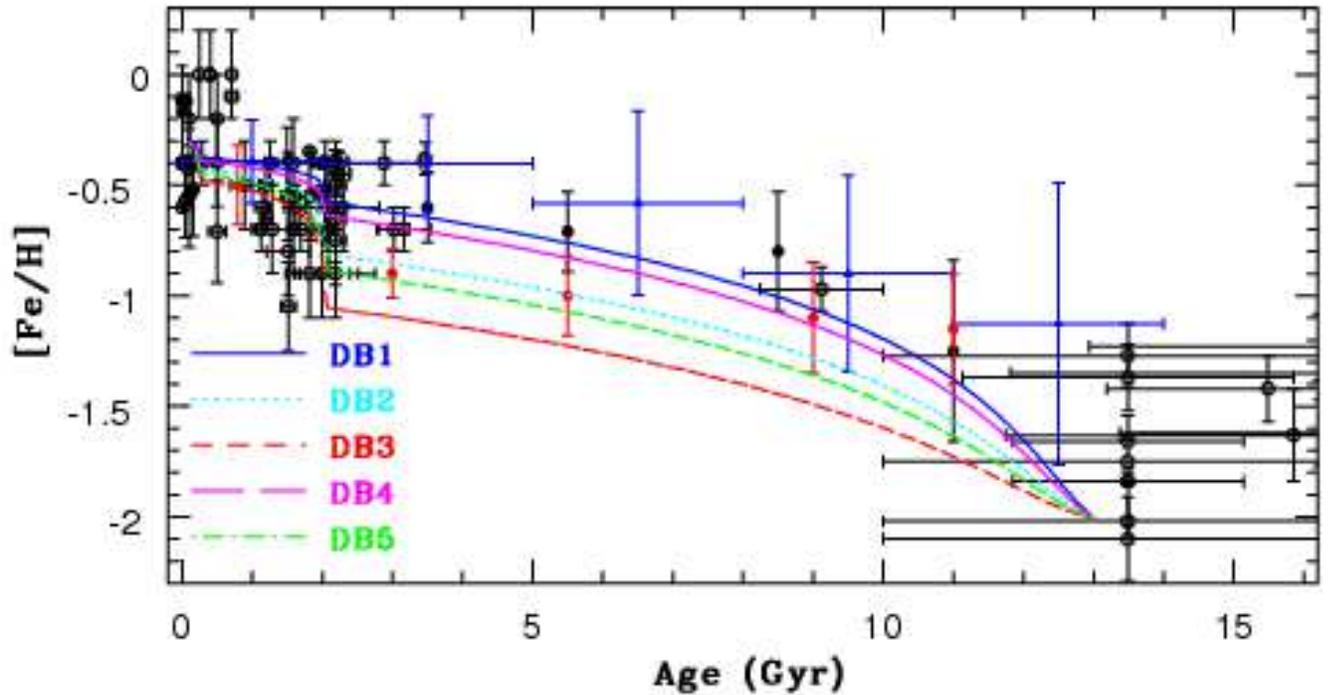}
\figcaption{
The same as Figure 3 but for the five double-burst models,
DB1 with
$\alpha=2.35$ and $C_{\rm q}=0.004$ (blue solid),
DB2 with
$\alpha=2.55$ and $C_{\rm q}=0.004$ (cyan dotted),
DB3 with
$\alpha=2.75$ and $C_{\rm q}=0.004$(red short-dashed),
DB4 with
$\alpha=2.55$ and $C_{\rm q}=0.006$ (magenta long-dashed),
and DB5 with
$\alpha=2.75$ and $C_{\rm q}=0.006$  (green dot-dashed).
The strength of the starbursts ($C_{\rm sb1}$ and $C_{\rm sb2}$) 
at 2 Gyr ago  and 0.2 Gyr ago
($t_{\rm sb1,s} \le t \le t_{\rm sb1,e}$ and
$t_{\rm sb2,s} \le t \le t_{\rm sb2,e}$) 
at each model are  chosen
such that the final [Fe/H] is the same as the observed one (i.e., $-0.3$).
The parameters $t_{\rm sb1, s}$ and $t_{\rm sb1,e}$ 
($t_{\rm sb2, s}$ and $t_{\rm sb2,e}$) can control the start and the end
of  the first (second) starburst, respectively, in the LMC.
\label{fig-11}}
\end{figure}

\begin{figure}
\plotone{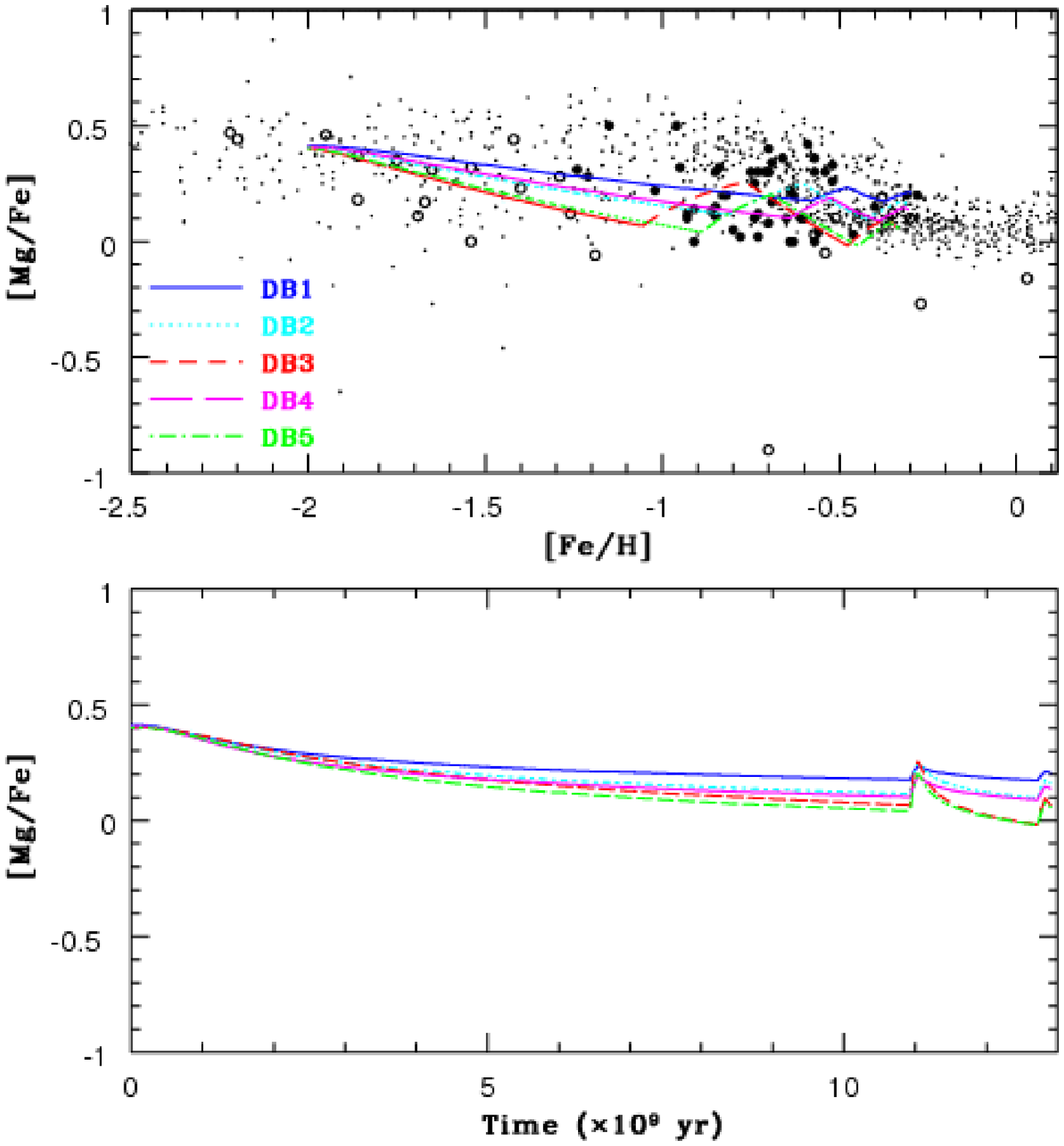}
\figcaption{
The same as Figure 4 but for the five double-burst models,
DB1 with
$\alpha=2.35$ and $C_{\rm q}=0.004$ (blue solid),
DB2 with
$\alpha=2.55$ and $C_{\rm q}=0.004$ (cyan dotted),
DB3 with
$\alpha=2.75$ and $C_{\rm q}=0.004$(red short-dashed),
DB4 with
$\alpha=2.55$ and $C_{\rm q}=0.006$ (magenta long-dashed),
and DB5 with
$\alpha=2.75$ and $C_{\rm q}=0.006$  (green dot-dashed).
\label{fig-12}}
\end{figure}

\begin{figure}
\plotone{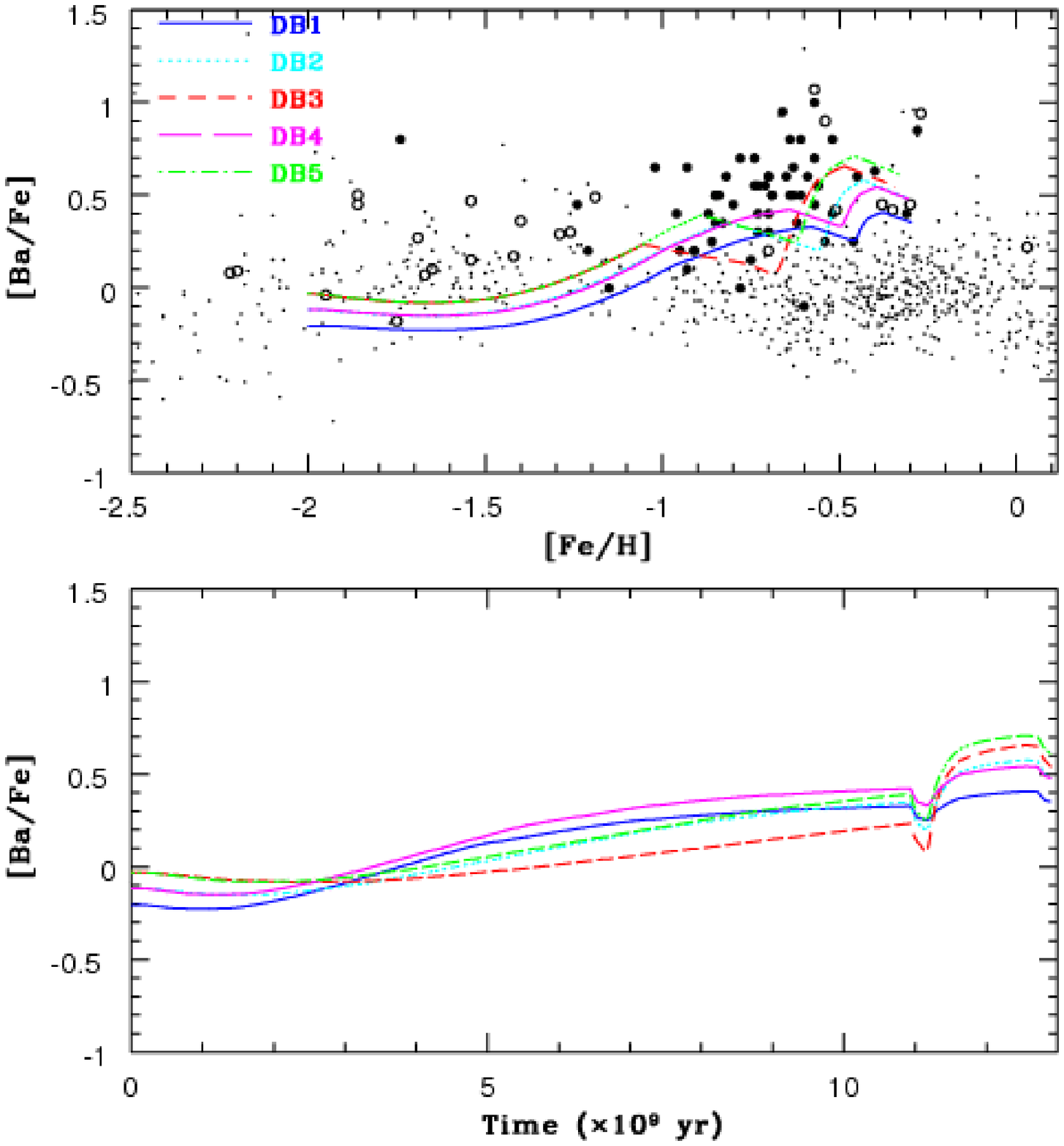}
\figcaption{
The same as Figure 5 but for the five double-burst models,
DB1 with
$\alpha=2.35$ and $C_{\rm q}=0.004$ (blue solid),
DB2 with
$\alpha=2.55$ and $C_{\rm q}=0.004$ (cyan dotted),
DB3 with
$\alpha=2.75$ and $C_{\rm q}=0.004$(red short-dashed),
DB4 with
$\alpha=2.55$ and $C_{\rm q}=0.006$ (magenta long-dashed),
and DB5 with
$\alpha=2.75$ and $C_{\rm q}=0.006$  (green dot-dashed).
\label{fig-13}}
\end{figure}

\begin{figure}
\plotone{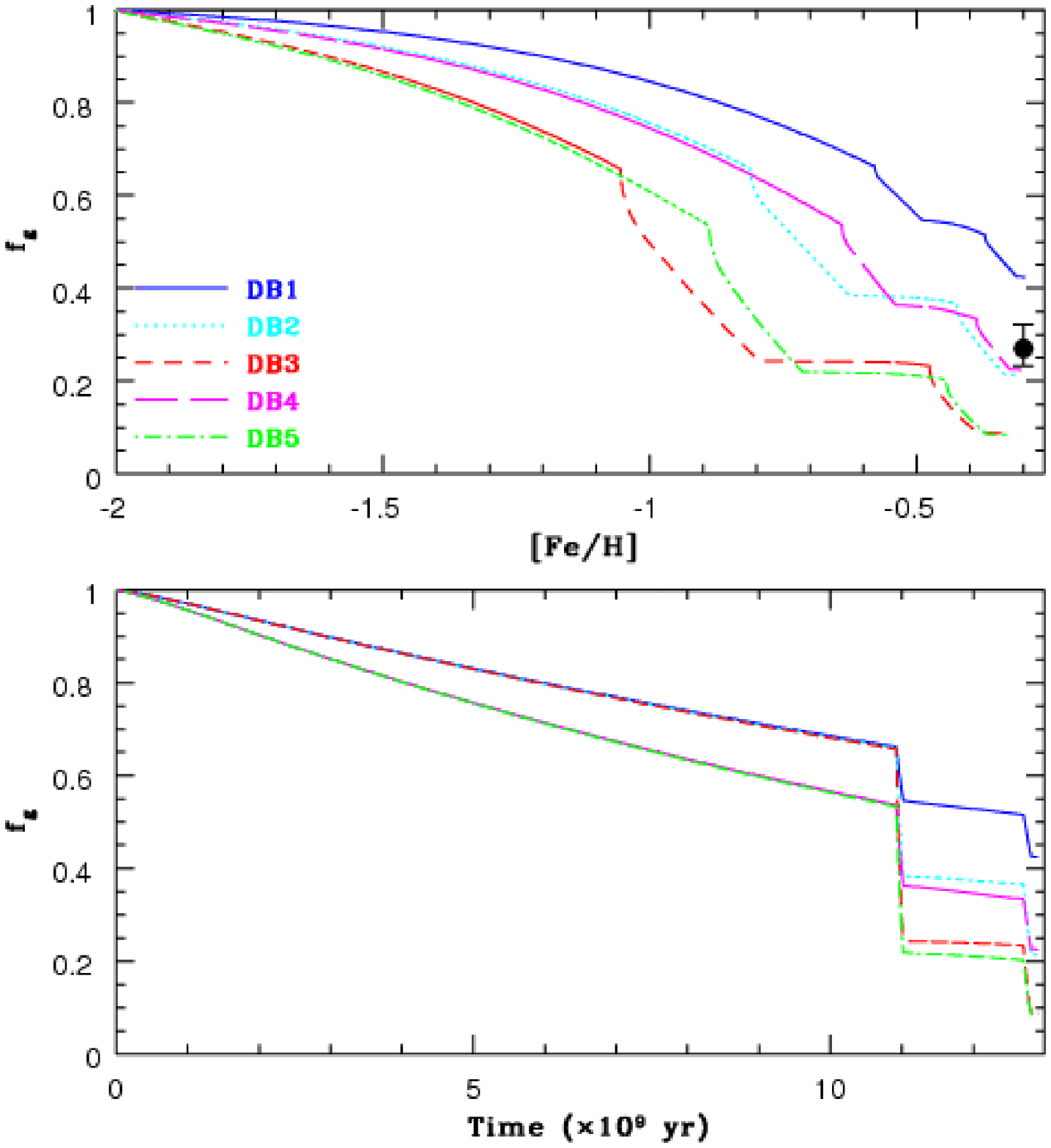}
\figcaption{
The same as Figure 6 but for the five double-burst models,
DB1 with
$\alpha=2.35$ and $C_{\rm q}=0.004$ (blue solid),
DB2 with
$\alpha=2.55$ and $C_{\rm q}=0.004$ (cyan dotted),
DB3 with
$\alpha=2.75$ and $C_{\rm q}=0.004$(red short-dashed),
DB4 with
$\alpha=2.55$ and $C_{\rm q}=0.006$ (magenta long-dashed),
and DB5 with
$\alpha=2.75$ and $C_{\rm q}=0.006$  (green dot-dashed).
\label{fig-14}}
\end{figure}

\begin{figure}
\plotone{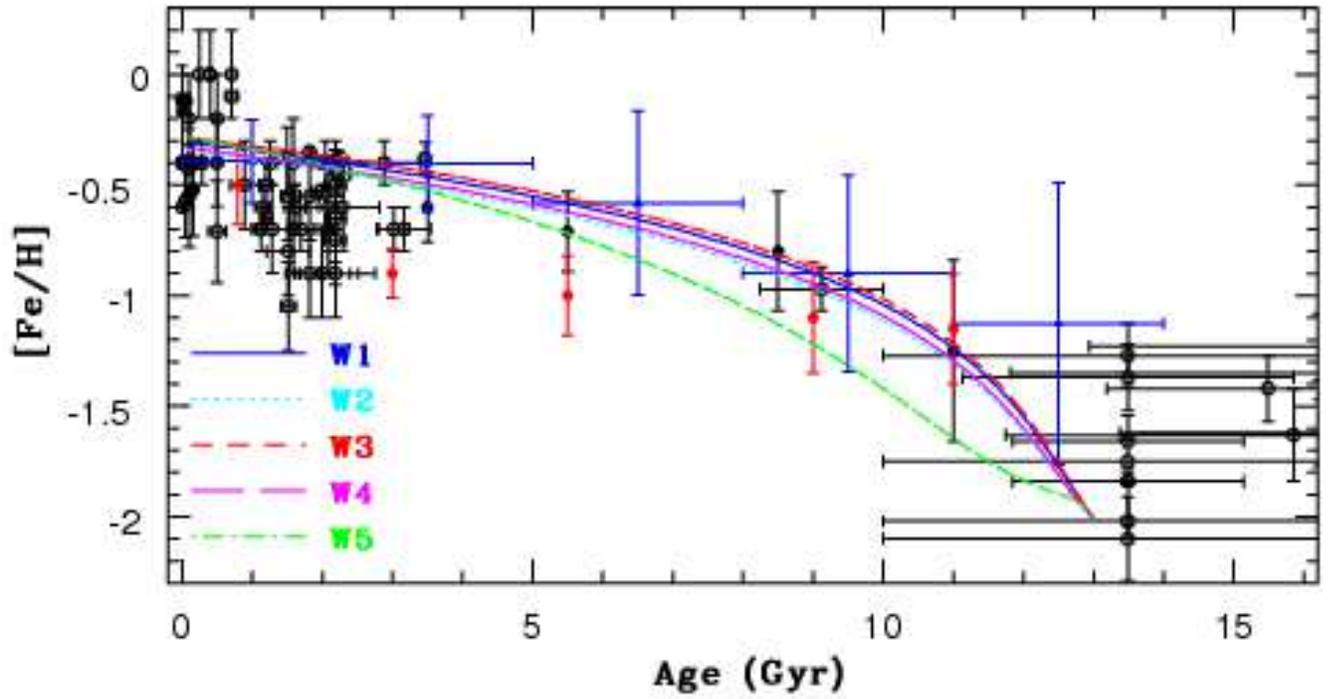}
\figcaption{
The same as Figure 3 but for the five wind models,
W1 with
$\alpha=2.35$ and $f_{\rm ej}=0.4$ (blue solid),
W2 with
$\alpha=2.55$ and $f_{\rm ej}=0.4$ (cyan dotted),
W3 with
$\alpha=2.35$ and $f_{\rm ej}=0.2$(red short-dashed),
W4 with
$\alpha=2.55$ and $f_{\rm ej}=0.2$ (magenta long-dashed),
and W5 with
$\alpha=2.35$ and $C_{\rm ej}=300$  (green dot-dashed).
C1--C4 are selective wind models  whereas C5 is a non-selective
wind model.  Starbursts do not  occur in these wind models. 
\label{fig-15}}
\end{figure}

\begin{figure}
\plotone{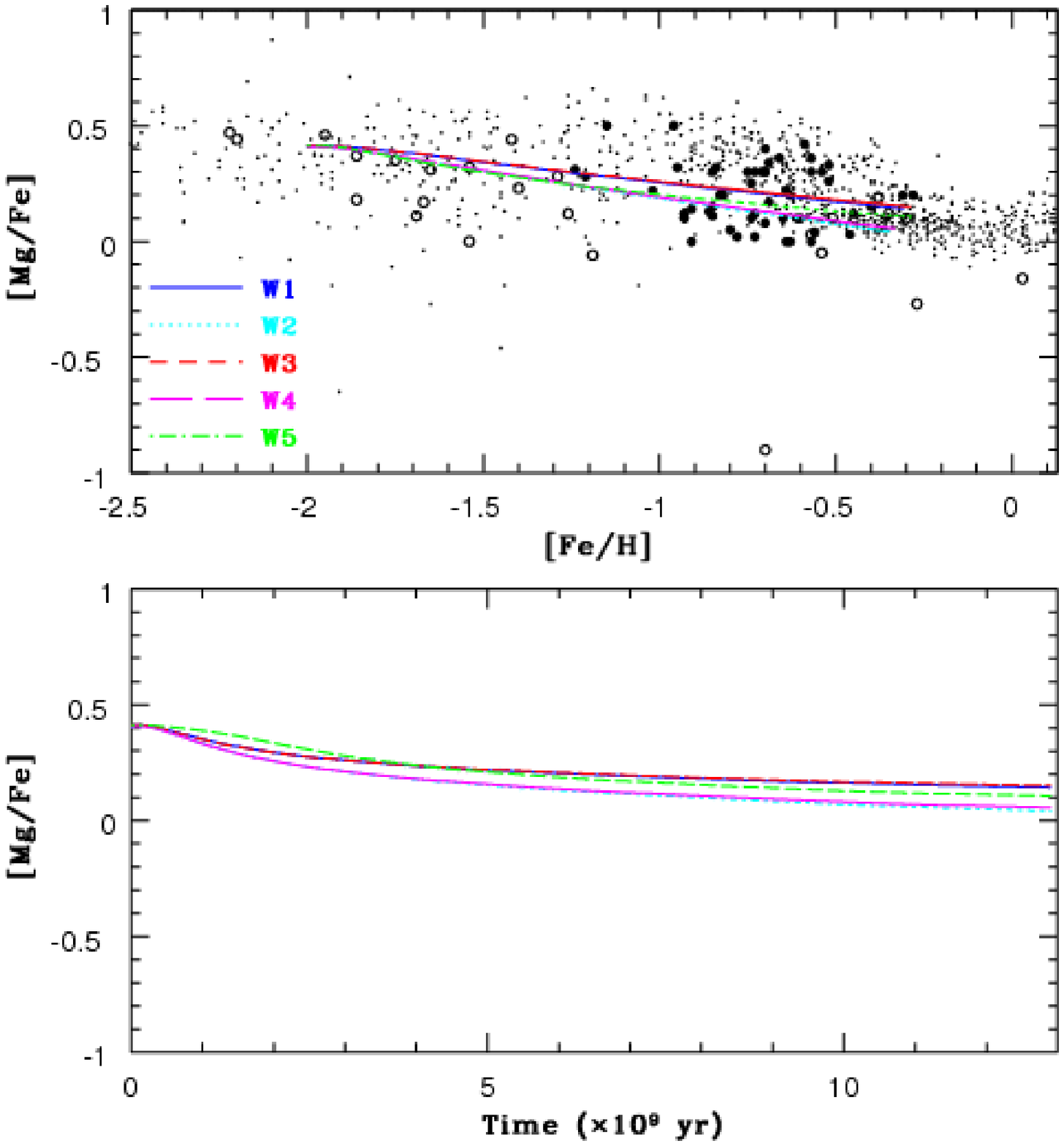}
\figcaption{
The same as Figure 4 but for the five wind models,
W1 with
$\alpha=2.35$ and $f_{\rm ej}=0.4$ (blue solid),
W2 with
$\alpha=2.55$ and $f_{\rm ej}=0.4$ (cyan dotted),
W3 with
$\alpha=2.35$ and $f_{\rm ej}=0.2$(red short-dashed),
W4 with
$\alpha=2.55$ and $f_{\rm ej}=0.2$ (magenta long-dashed),
and W5 with
$\alpha=2.35$ and $C_{\rm ej}=300$  (green dot-dashed).
\label{fig-16}}
\end{figure}

\begin{figure}
\plotone{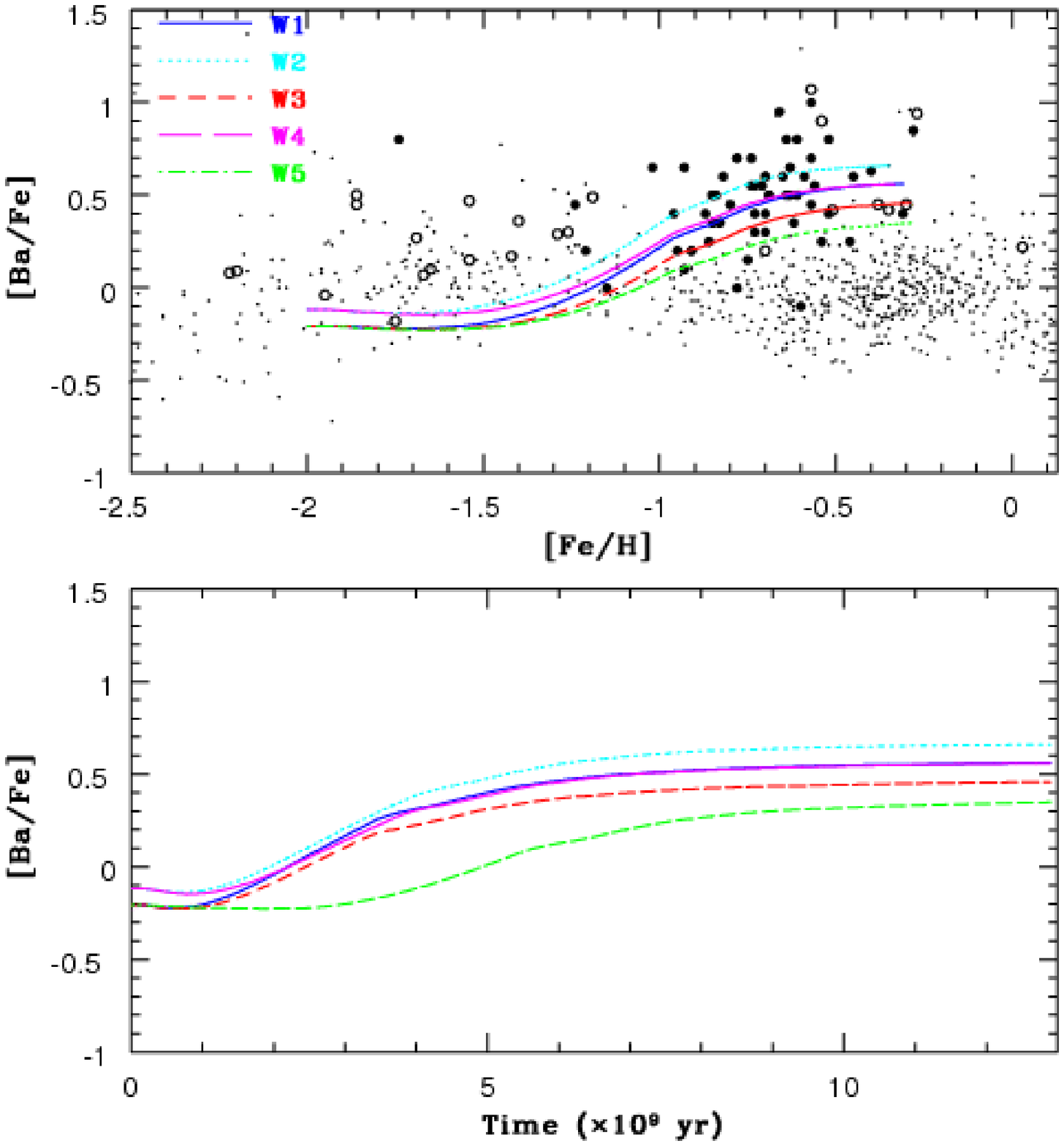}
\figcaption{
The same as Figure 5 but for the five wind models,
W1 with
$\alpha=2.35$ and $f_{\rm ej}=0.4$ (blue solid),
W2 with
$\alpha=2.55$ and $f_{\rm ej}=0.4$ (cyan dotted),
W3 with
$\alpha=2.35$ and $f_{\rm ej}=0.2$(red short-dashed),
W4 with
$\alpha=2.55$ and $f_{\rm ej}=0.2$ (magenta long-dashed),
and W5 with
$\alpha=2.35$ and $C_{\rm ej}=300$  (green dot-dashed).
\label{fig-17}}
\end{figure}

\begin{figure}
\plotone{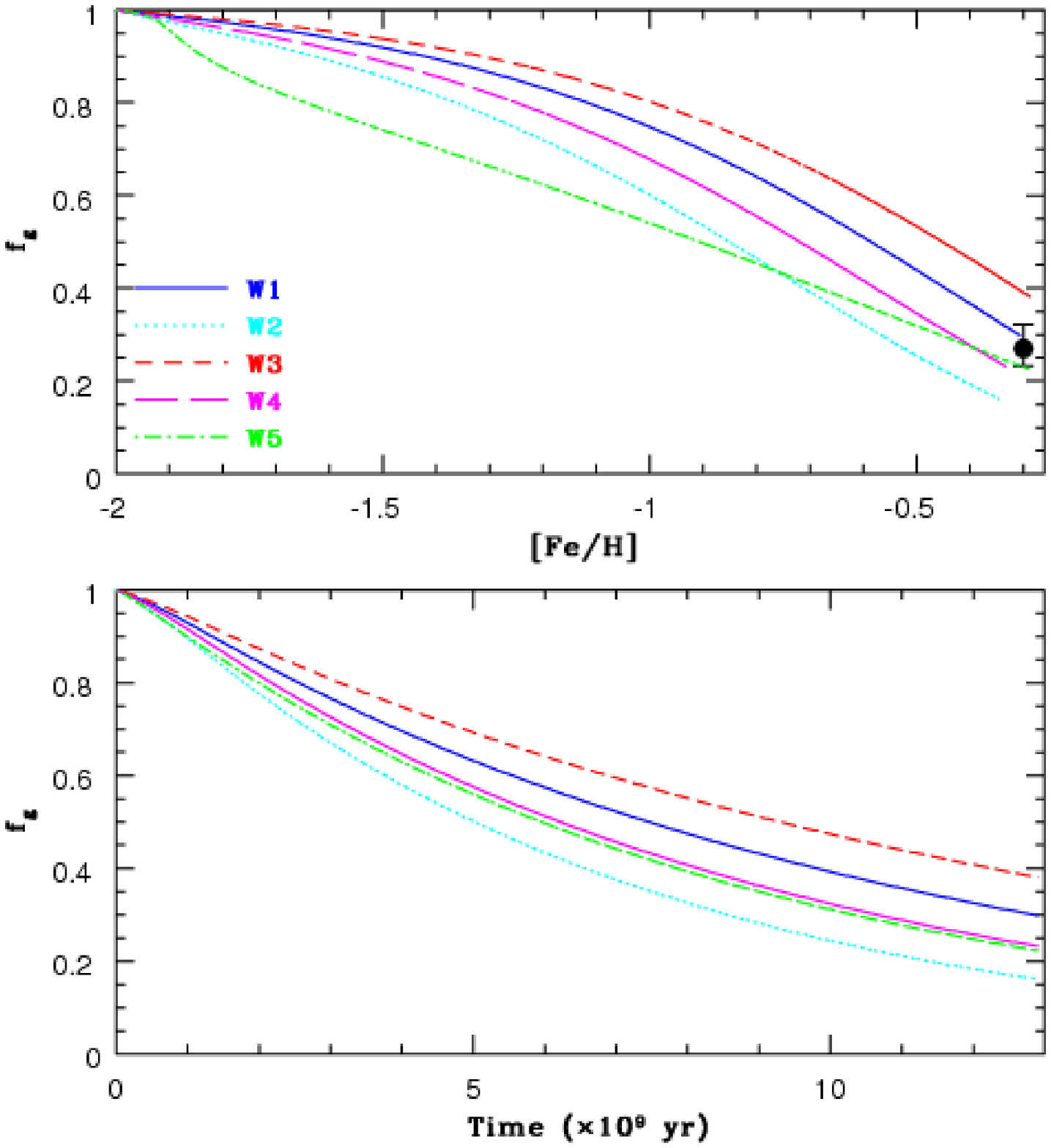}
\figcaption{
The same as Figure 6 but for the five wind models,
W1 with
$\alpha=2.35$ and $f_{\rm ej}=0.4$ (blue solid),
W2 with
$\alpha=2.55$ and $f_{\rm ej}=0.4$ (cyan dotted),
W3 with
$\alpha=2.35$ and $f_{\rm ej}=0.2$(red short-dashed),
W4 with
$\alpha=2.55$ and $f_{\rm ej}=0.2$ (magenta long-dashed),
and W5 with
$\alpha=2.35$ and $C_{\rm ej}=300$  (green dot-dashed).
\label{fig-18}}
\end{figure}

\clearpage

\begin{figure}
\plotone{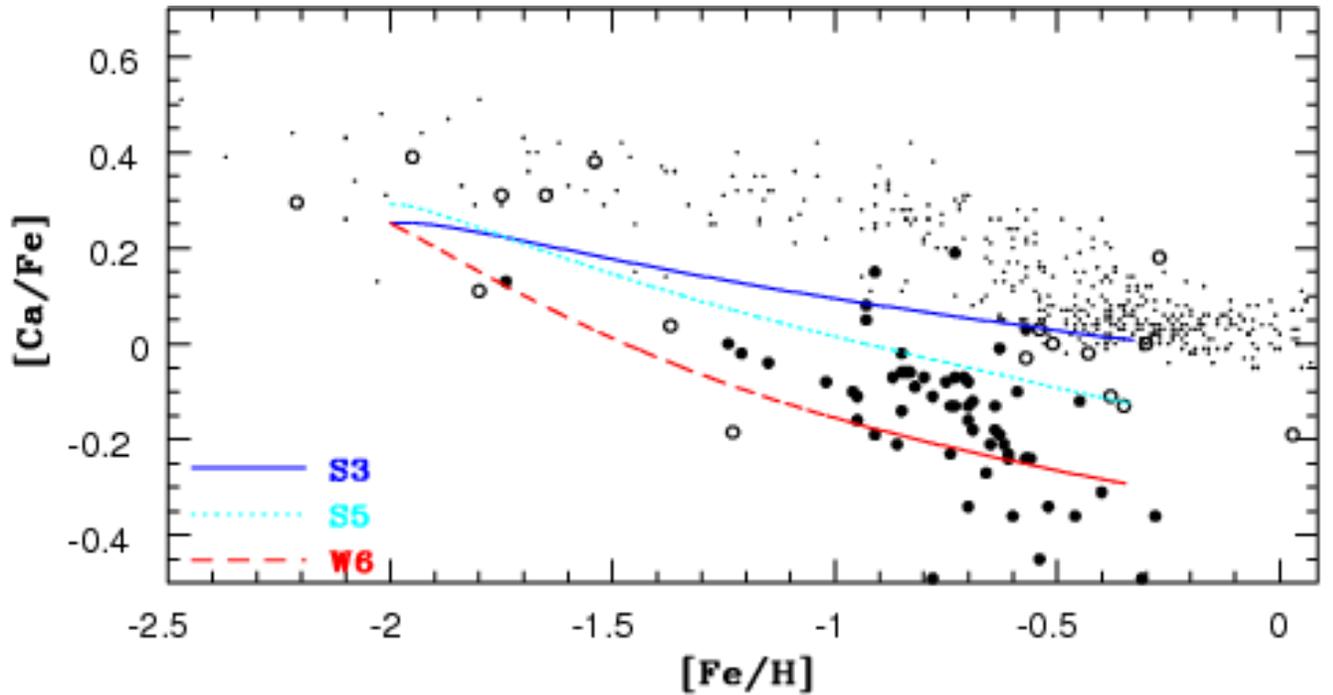}
\figcaption{
Chemical evolution on the [Ca/Fe]-[Fe/H] plane for the three
models S3 (blue solid), S5 (cyan dotted), and W6 (red dashed).
The observational plots are from observational studies shown
in Figure 4.  The wind model W6 is different from other wind models
in that Ca can be by a factor of 1.7 more efficiently removed from
the LMC disk in comparison with other elements from SNe. This selective
removal of Ca is based on previous theoretical studies on nucleosynthesis
of jet-induced SNe (Shigeyama  et al. 2010). Note that the wind model W6
can be better fit to the observed low [Ca/Fe] ($<-0.2$) for the LMC
field stars with [Fe/H]$>-1.0$.
\label{fig-19}}
\end{figure}

\newpage

\begin{figure}
\plotone{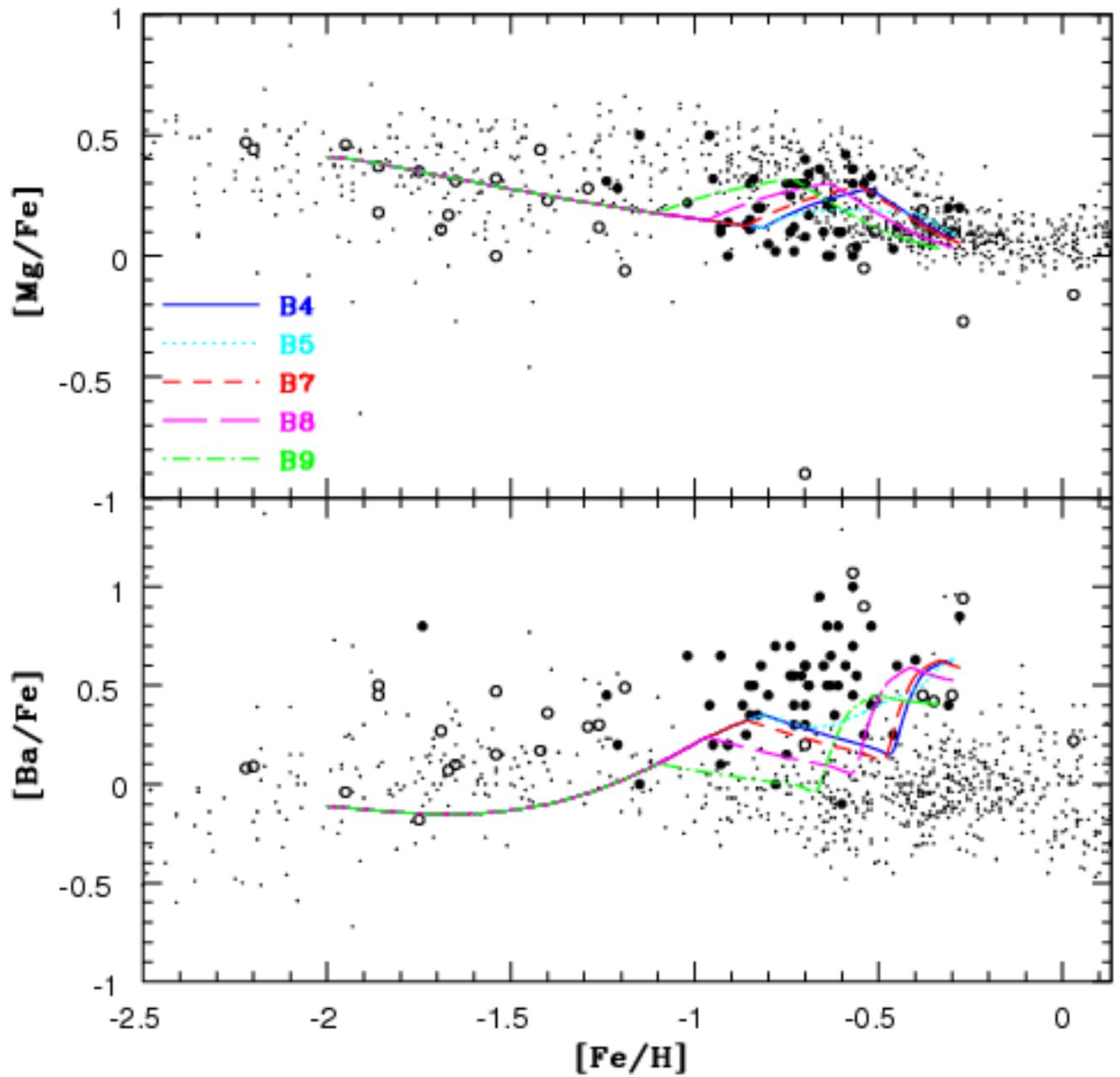}
\figcaption{
Chemical evolution on the [Mg/Fe]-[Fe/H] plane (upper)
and the [Ba/Fe]-[Fe/H] one (lower) for the five starburst models 
with different epoch and duration of starburst;
B4
(blue solid),
B5 
(cyan dotted),
B7
(red short-dashed),
B8 
(magenta long-dashed),
and  B9
(green dot-dashed).
\label{fig-20}}
\end{figure}

\clearpage

\begin{figure}
\plotone{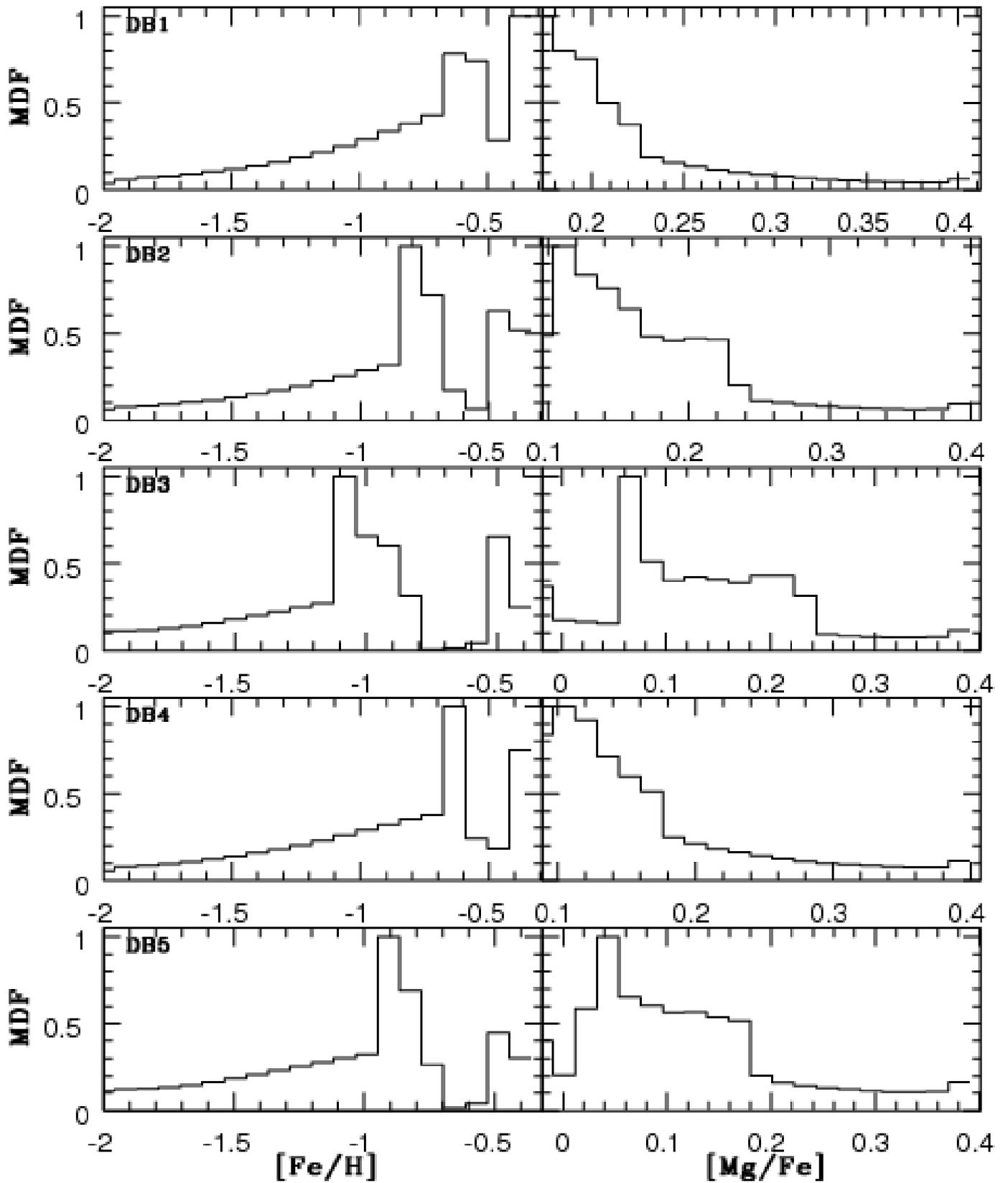}
\figcaption{
The MDFs for [Fe/H] (left)
and [Mg/Fe] (right) in the five double-burst models, DB1 (top),
DB2 (second from the top),
DB3 (middle),
DB4 (second from the bottom),
and DB5 (bottom). The MDFs are normalized to the maximum numbers of
stars in the bins. 
\label{fig-21}}
\end{figure}

\clearpage

\begin{figure}
\plotone{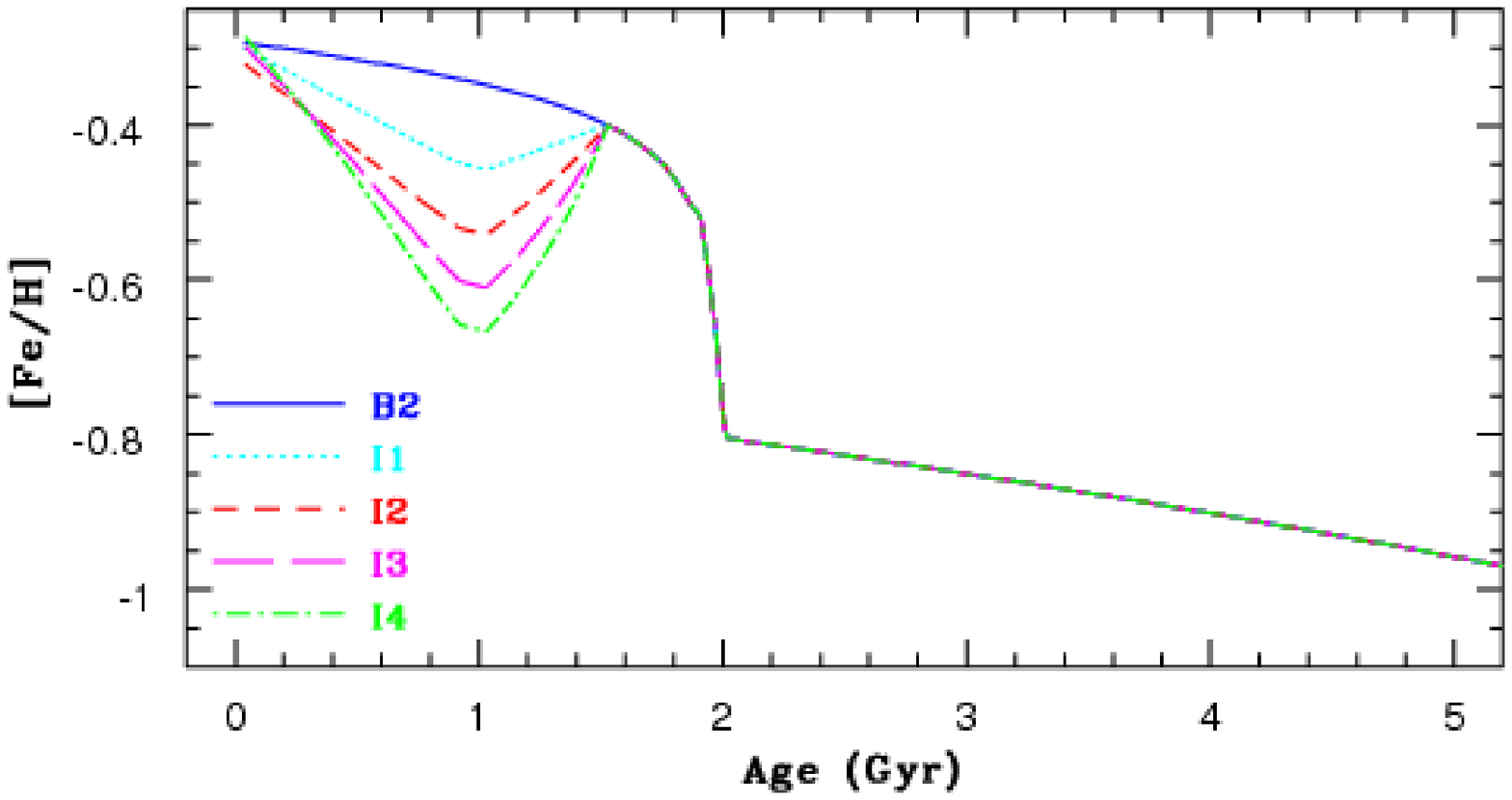}
\figcaption{
The AMRs for the last $\sim 5$ Gyr for the burst model B2 (blue solid)
and the four infall models,
I1 with $M_{\rm ext}=0.1$ 
(cyan dotted),
I2 with $M_{\rm ext}=0.2$ 
(red short-dashed),
I3 with $M_{\rm ext}=0.3$ 
(magenta long-dashed),
and I4 with $M_{\rm ext}=0.4$ 
(green dot-dashed).
Here $M_{\rm ext}$ represents
the total mass of external  metal-poor gas accreted onto the LMC.
$M_{\rm ext}=1$ thus means that the total amount of external gas accretion
is the same as the total amount of gas accreted onto the  LMC
from  its own halo for the last 13 Gyr.
Infall of external metal-poor gas onto the LMC disk
is assumed to commence  1.5 Gyr  ago and end
1 Gyr ago  in these four infall models.
Starbursts are assumed to occur twice at 2 Gyr ago and 1 Gyr
(just after the end of the external gas infall event) in these infall models
so that the final stellar metallicity in each model
can be the same as the observed one
([Fe/H]$\sim -0.3$). 
\label{fig-22}}
\end{figure}


\begin{thebibliography}{}
\bibitem[]{}
Ardeberg, A., Gustafsson, B.,  Linde, P., \&  Nissen, P.-E.
1997, A\&A, 322, L13
\bibitem[)]{}
Bekki, K. 2011, MNRAS, 412, 2241
\bibitem[)]{}
Bekki, K., \& Chiba, M. 
2005, MNRAS, 356, 680
\bibitem[Bekki \& Chiba(2007)]{Bekki_07}
Bekki, K., \& Chiba, M. 2007, MNRAS, 381, L16 (BC07)
\bibitem[]{}
Bekki, K. Campbell, S. W. Lattanzio, J. C., \&  Norris, J. E.
2007, MNRAS, 377, 335
\bibitem[)]{}
Bekki, K.,  \& Sanimirovic, S. 2009, MNRAS, 395, 342
\bibitem[)]{}
Bekki, K.,  \& Tsujimoto, T., 2010, ApJ, 721, 1515
\bibitem[)]{}
Bell, E. F., \& de Jong, R. S. 2001, ApJ, 550, 212
\bibitem[Bensby et al.(2005)]{Bensby_05}
Bensby, T., Feltzing, S., Lundstr\"{o}m, I., \& Ilyin, I. 2005, A\&A, 433, 185
\bibitem[]{}
Bernard, J.-P., et al. 2008, AJ, 136 919
\bibitem[]{}
Bertelli, G., Mateo, M. Chiosi, C., \&  Bressan, A.
1992, ApJ, 388, 400
\bibitem[]{}
Bica, E., Dottori, H., \&  Pastoriza, M.	
1986, A\&A, 156, 261
\bibitem[Busso et al.(2001)]{Busso_01}
Busso, M., Gallino, R., Lambert, D. L., 
Travaglio, C., \& Smith, V. V. 2001, ApJ, 557, 802
\bibitem[]{}
Butcher, H. 
1977, ApJ, 216, 372
\bibitem[]{}
Carrera, R., Gallart, C., Hardy, E.,  Aparicio, A., \&  Zinn, R.
2008, AJ, 135, 836 (C08) 
\bibitem[]{}
Cioni, M.-R. L.,  Girardi, L.,  Marigo, P.,  \& Habing, H. J.
2006, A\&A, 448, 77
\bibitem[Colucci et al.(2012)]{Colucci_12}
Colucci, J. E., Bernstein, R. A., Cameron, S. A., \& McWilliam, A.
 2012, ApJ, 746, 29 (C12) 
\bibitem[]{}
Cole, A. A., Tolstoy, E.,  Gallagher, J. S. III.,  \& Smecker-Hane, T.  A.
2005, AJ, 129, 1465 (C05)
\bibitem[]{}
Da Costa G. S. 1991, 
in Haynes R., Milne D., eds, Proc. IAU Symp. 148,
The Magellanic Clouds, Kluwer, Dordrecht, 
p183
\bibitem[Diaz \& Bekki(2012)]{Diaz_12}
Diaz, J. D., \& Bekki, K. 2012, ApJ, in press, arXiv1112.6191,  (DB12)
\bibitem[]{}
Dopita, M. A., et al.
1997, ApJ, 474, 188
\bibitem[]{}
Elson R. A. W., Gilmore, G. F., \&  Santiago, B. X.
1997, MNRAS,  289, 157
\bibitem[]{}
Gallagher et al. 
1996, ApJ, 466, 732
\bibitem[Gallino et al.(1998)]{Gallino_98}
Gallino, R., et al. 1998, ApJ, 497, 388
\bibitem[]{}
Geha, M. C., et al. 1998, AJ, 115, 1045
\bibitem[]{}
Geisler, D.,  Bica, E., Dottori, H., Claria, J. J.,
Piatti, A. E., \&  Santos, J. F. C., Jr. 1997,
AJ, 114, 1920 
\bibitem[]{}
Geisler, D.,  Piatti, A.  E.,  Bica, E.,  Clari\'a, J. J.
2005, MNRAS, 341, 771
\bibitem[]{}
Glatt, K., Grebel, E. K., \& Koch, A. 2010, A\&A, 517, 50
\bibitem[Gratton et al.(2003)]{Gratton_03}
Gratton, R. G., Carretta, E., Claudi, R., Lucatello, S., \& Barbieri, M. 2003, A\&A, 404, 187
\bibitem[]{} 
Harris J., \& Zaritsky D., 2009, AJ, 138, 1243 (HZ09)
\bibitem[]{} 
Hascheke, R., et al. 2012 in preprint (arXiv:1207.5791) 
\bibitem[]{}
Hill, V.,  Andrievsky, S., \&  Spite, M.
1995, A\&, 293, 347
\bibitem[]{}
Hill, V.,  Fran\c ois, P., Spite, M., Primas, F., \&  Spite, F.
2000, A\&A, 364, L19
\bibitem[]{}
Holtzman, J. A., et al. 1997, AJ, 113, 656
\bibitem[]{}
Holtzman et al. 1998, AJ, 115, 1946
\bibitem[]{}
Indu, G.,  \& Subramaniam, A. 2011, A\&A, 535, 115
\bibitem[]{}
Hill, R. J., Madore, B. F.,  \& Freedman, W. L. 1994, ApJ, 429, 204
\bibitem[Ishimaru \& Wanajo(1999)]{Ishimaru_99}
Ishimaru, Y., \& Wanajo, S. 1999, ApJ, 511, L33
\bibitem[Johnson et al.(2006)]{Johnson_06}
Johnson, J. A., Ivans, I. I., \& Stetson, P. B. 2006, ApJ, 640, 801
\bibitem[Kim et al.(1999)]{Kim_99}
Kim, S., Dopita, M. A., Staveley-Smith, L., \& Bessel, M. 1999, AJ, 118, 2823
\bibitem[]{}
Luck, R.  E.,  Moffett, T. J., Barnes, T. G. III., \& Gieren, W. P.
1998, AJ, 115, 605
\bibitem[Mannucci et al.(2006)]{Mannucci_06}
Mannucci, F., Della Valle, M., \& Panagia, N. 2006, MNRAS, 370, 773
\bibitem[Maoz et al.(2010)]{Maoz_10}
Maoz, D., Sharon, K., \& Gal-Yam, A. 2010, ApJ, 722, 1879
\bibitem[]{}
Maoz, D., Mannucci, F.,  Li, W.,  Filippenko, A. V.,
Della Valle, M.,  \&  Panagia, N.
2011, MNRAS, 412, 1508
\bibitem[]{}
Maoz, D., \& Badenes, C. 2010, MNRAS, 407, 1314
\bibitem[]{}
Mateo, M. 1988, ApJ, 331, 261
\bibitem[Mathews et al.(1992)]{Mathews_92}
Mathews, G. J., Bazan, G., \& Cowan, J. J. 1992, ApJ, 391, 719
\bibitem[Mucciarelli et al.(2008)]{Mucciarelli_08}
Mucciarelli, A., Carretta, E., Origlia, L., \& Ferraro, F. R. 
2008, AJ, 136, 375 
\bibitem[Mucciarelli et al.(2010)]{Mucciarelli_10}
Mucciarelli, A., Origlia, L., \& Ferraro, F. R. 2010, ApJ, 717, 277 (M10)
\bibitem[Mucciarelli et al.(2011)]{Mucciarelli_11}
Mucciarelli, A., et al. 2011, MNRAS, 413, 837 (M11)
\bibitem[]{}
Olszewski, E. W., Schommer, R.  A., Suntzeff, N. B., \& Harris, H. C.
1991, AJ, 101, 515
\bibitem[]{}
Olsen, K. A. G. 1999, AJ, 117, 2244
\bibitem[]{}
Olsen, K. A. G., Zaritsky, D.,  Blum, R.  D., Boyer, M. L., \& Gordon, K. D.
2011, ApJ, 737, 29
\bibitem[]{}
Olszewski, E. W., Suntzeff, N. B. \&  Mateo, M.
1996, ARA\&A, 34, 511
\bibitem[]{}
Piatti, A. E. 2011, MNRAS, 418, L40
\bibitem[Pagel \& Tautvai\v{s}ien\'{e}(1998)]{Pagel_98}
Pagel, B. E. J., \& Tautvai\v{s}ien\'{e}, G. 1998, MNRAS, 299, 535 (PT98)
\bibitem[]{}
Pilyugin, L. S. 1996, A\&A, 313, 803
\bibitem[Pomp\'{e}ia et al.(2008)]{Pompeia_08}
Pomp\'{e}ia, L., et al. 2008, A\&A, 480, 379 (P08)
\bibitem[Reddy et al.(2003)]{Reddy_03}
Reddy, B. E., Tomkin, J., Lambert, D. L., \& Allende Prieto, C. 2003, MNRAS, 340, 304
\bibitem[]{}
Renzini, A., \&  Buzzoni, A., 1986,
in  Spectral Evolution of Galaxies,
Reidel, Dordrecht, p.195
\bibitem[]{}
Rubele, S., et al., 2012, A\&A, 537, 106
\bibitem[]{}
Russell, S. C., \&  Dopita, M.  A. 
1992, ApJ, 384, 508
\bibitem[]{}
Salpeter, E. E. 1955, ApJ, 121, 161
\bibitem[]{}
Shigeyama, T., et al. 2010, AIPC, 1279, 415
\bibitem[]{}
Smecker-Hane, T. A., Cole, A. A., Gallagher, J.  S. III,
\& Stetson, P. B.
2002, ApJ, 566, 239.
\bibitem[]{}
Stryker, L. 1983, ApJ, 266, 82
\bibitem[]{}
Subramaniam, A.,  \& Prabhu, T.  P. 2005, ApJ, L47
\bibitem[Totani et al.(2008)]{Totani_08}
Totani, T., Morokuma, T., Oda, T., Doi, M., \& Yasuda, N. 2008, PASJ, 60, 1327
\bibitem[Tsujimoto(2012)]{Tsujimoto_12}
Tsujimoto, T. 2012, ApJ, 736, 113
\bibitem[]{}
Tsujimoto, T., Nomoto, K., Yoshii, Y., Hashimoto, M., Yanagida, S.,
\& Thielemann, F.-K. 1995, MNRAS, 277, 945 (T95)
\bibitem[]{}
Tsujimoto, T.,  Bland-Hawthorn, J.,  \& Freeman, K. C. 2010,  PASJ, 62, 447
\bibitem[]{}
Tsujimoto, T., \&  Bekki, K. 2012, ApJ, 747, 125 (TB12)
\bibitem[]{}
Vallenari, A., Chiosi, C., Bertelli, G., Aparicio, A., \&  Ortolani, S.
1996, A\&A, 309, 367
\bibitem[]{}
Weidemann, V. 2000, A\&A, 363, 647
\bibitem[Wheeler et al.(1998)]{Wheeler_98}
Wheeler, J. C., Cowan, J. J., \& Hillebrandt, W. 1998, ApJ, 493, 101
\bibitem[Yoshii et al.(1996)]{Yoshii_96}
Yoshii, Y., Tsujimoto, T., \& Nomoto, K. 1996, ApJ, 462, 266
\end{thebibliography}
\end{document}